\documentclass[a4paper]{article}

\usepackage[english]{babel}
\usepackage[utf8x]{inputenc}
\usepackage[T1]{fontenc}
\usepackage{ upgreek }
\usepackage{gensymb} 
\usepackage{threeparttable}
\usepackage{tabularx}
\usepackage{ntheorem}
\newtheorem{theorem}{Theorem}

\usepackage{cite}
\usepackage{mathrsfs}
\usepackage{pifont}
\usepackage{amssymb}
\usepackage{array, caption}
\usepackage{amsmath}
\usepackage{booktabs}
\usepackage{graphicx}
\usepackage{subcaption}
\makeatletter
\newcommand{\ssymbol}[1]{^{\@fnsymbol{#1}}}
\makeatother

\usepackage{amsmath}
\usepackage{graphicx}
\usepackage[colorinlistoftodos]{todonotes}
\usepackage[colorlinks=true, allcolors=blue]{hyperref}

\usepackage[a4paper,top=2.5cm,bottom=2cm,left=2.5cm,right=2.5cm,marginparwidth=2cm]{geometry}

\usepackage{authblk}
\title{ A Feasible Semi-quantum Private Comparison Based on Entanglement Swapping of Bell States}

\author[1]{Chong-Qiang Ye}
\author[2]{Jian Li \thanks{Corredponding author:  lijian@bupt.edu.cn}}
\author[2]{Xiu-Bo Chen}
\author[1]{Yanyan Hou}

\affil[1]{ School of Artificial Intelligence, Beijing University of Posts Telecommunications, Beijing 100876, China.}
\affil[2]{Information Security Center, State Key Laboratory of Networking and Switching Technology,
Beijing University of Posts and Telecommunications, Beijing 100876, China}

\begin{document}
\date{}
 \maketitle

\begin{abstract}
Semi-quantum private comparison (SQPC) enables two classical users with limited quantum capabilities to compare confidential information using a semi-honest third party (TP) with full quantum power. However, entanglement swapping, as an important property of quantum mechanics in previously proposed SQPC protocols is usually neglected. In this paper, we propose a feasible SQPC protocol based on the entanglement swapping of Bell states, where two classical users do not require additional implementation of the semi-quantum key distribution protocol to ensure the security of their private data. Security analysis shows that our protocol is resilient to both external and internal attacks. To verify the feasibility and correctness of the proposed SQPC protocol, we design and simulate the corresponding quantum circuits using IBM Qiskit. Finally, we compare and discuss the proposed protocol with previous similar work. The results reveal that our protocol maintains high qubit efficiency, even when entanglement swapping is employed. Consequently, our proposed approach showcases the potential applications of entanglement swapping in the field of semi-quantum cryptography.

\end{abstract}

\section{Introduction}
\label{intro}

Secure multi-party computation (SMC) \cite{1,2,3} is an essential branch in classical cryptography, which allows several users to jointly compute a function based on their secret data while maintaining data privacy. SMC has a wide range of applications in scenarios such as secret ballot elections, private comparison, and auctions. However, with the vigorous development of quantum technology, the security of SMC protocols based on computational complexity has been threatened by adversaries with quantum computing power \cite{4,5}. In order to meet this challenge, the quantum secure multi-party computation (QSMC) was proposed, which uses quantum technology to protect the data privacy of participants and the security of calculation results\cite{6}. Due to its unique security, QSMC has received extensive attention and research. Many QSMC protocols have been proposed, including quantum summation \cite{7,8,9}, quantum private query \cite{10,11,12} and quantum private comparison \cite{13,14,15,16,17,18}.

Quantum private comparison (QPC)  is a significant field within QSMC.  Its objective is to address the comparison of private data equality among multiple users while ensuring that their respective data is not revealed. Generally, a QPC protocol needs to meet the following conditions: (1) Fairness: All users should obtain the comparison result simultaneously. (2) Security: The data of each user is confidential and unavailable to both the other users and the third party (conventionally called TP) \cite{19}.  For further investigation on this topic, the reader is referred to the literature cited in reference \cite{20}.

Traditionally, QPC requires all users to possess full quantum capabilities to ensure the security of the protocol. However, given the current technological limitations, not all users may be able to afford the high costs associated with quantum devices. To overcome this issue, Boyer et al. \cite{21,22} proposed a novel concept of semi-quantum. In the semi-quantum settings, only one user possesses full quantum capabilities, typically referred to as the ``quantum'' user, while the remaining users are restricted to performing simple quantum operations and are often referred to as ``classical'' or ``semi-quantum'' users. The classical users can usually only do the following operations: (a) measuring the qubits in $Z$ basis $\{|0\rangle, |1\rangle\}$; (b) preparing the qubits in $Z$ basis; (c) sending or reflecting the qubits without disturbance.
Semi-quantum provides an effective solution to alleviate the current shortage of quantum resources, and the research on semi-quantum cryptography has attracted widespread attention.

Incorporating the concept of ``semi-quantum'' into QPC, Chou et al. \cite{23} constructed the first semi-quantum private comparison (SQPC) protocol, where two classical users, Alice and Bob, want to know whether their secret data are the same with the help of TP, who has full quantum capability and may be adversarial. This asymmetric setting of participant capabilities has made SQPC protocols particularly intriguing and garnered researchers' attention. From then on, scholars focused on SQPC and proposed many protocols based on different quantum states. For example, Ye et al. \cite{24} proposed a measure-resend SQPC protocol based on product states. In 2019, Lin et al. \cite{25} presented an SQPC protocol based on single photons, which eliminates the need for pre-shared keys between the classical users. Then, Thapliyala et al. \cite{26} and Jiang \cite{27} respectively designed an SQPC protocol using Bell states, where the pre-shared keys are required. In 2021, Yan et al. \cite{28} put forward an SQPC protocol using three-particle G-like states, in which pre-shared keys are also required. In the same year, Ye et al.\cite{29} proposed an SQPC protocol based on circular transmission. Unlike the previous protocols, in this protocol, the qubits travel from TP to Alice, to Bob, then back to TP. In 2022, Tian et al.\cite{30} utilized W-state as the information carrier to design SQPC protocol. In addition, several SQPC protocols based on $d$-level systems have been proposed \cite{31,32,33}. These protocols extend the scope of SQPC to encompass semi-quantum system with higher dimensions.

Despite progress in SQPC protocols, there remain limitations that require attention. Many existing protocols rely on additional semi-quantum key distribution protocols to construct pre-shared keys, ensuring private data security between classical users. However, using pre-shared keys often introduces additional qubit consumption costs, which in turn can increase the burden on the protocol's quantum resources. This presents a challenge for designing SQPC protocols that can be used in quantum resource-scarce scenarios. Additionally, previous protocols have not effectively utilized the entanglement swapping property due to the asymmetry of quantum capabilities between different users. While some have implemented entanglement swapping in SQPC protocols, the qubit efficiency is much lower than in protocols that do not use entanglement swapping \cite{23}. Nevertheless, entanglement swapping is an essential quantum feature, and it is worth exploring how it can be better utilized in the design of SQPC protocols.

In this work, we present a feasible SQPC protocol based on entanglement swapping of Bell states, utilizing Bell states as the information carrier and circular transmission to achieve information exchange between different users. The proposed protocol not only ensures the correctness and feasibility of secret data comparisons but also guarantees the security and privacy of the data. Compared to the previous SQPC protocols, our protocol has high qubit efficiency, which is an urgent requirement for current semi-quantum protocols.
Our contributions in this work are summarized as follows.

(1) We propose an efficient SQPC protocol for comparing the private data of two classical users by introducing entanglement swapping with Bell states and circular transmission. Even with the use of entanglement swapping, our protocol maintains a high level of efficiency.

(2) Our protocol does not require additional semi-quantum key distribution protocols for pre-shared key establishment to ensure the security of private data, thereby greatly reducing the protocol's complexity compared to those requiring pre-shared keys.

(3) IBM Qiskit is utilized to conduct the corresponding circuit simulations, validating the feasibility and correctness of the proposed SQPC protocol.


\section{Semi-quantum private comparison protocol}
\subsection{Preliminaries}
In this section, we describe the specific steps of the proposed SQPC protocol. First, we introduce the quantum resources used in this protocol, namely Bell states and entanglement swapping. In this paper, Bell states are defined as:
\begin{equation}
\begin{split}
\label{Bell-sequence}
|\phi^+\rangle=\frac{1}{\sqrt 2}(|00\rangle+|11\rangle), \quad |\phi^-\rangle=\frac{1}{\sqrt 2}(|00\rangle-|11\rangle),\\
|\psi^+\rangle=\frac{1}{\sqrt 2}(|01\rangle+|10\rangle), \quad |\psi^-\rangle=\frac{1}{\sqrt 2}(|01\rangle-|10\rangle).\\
\end{split}
\end{equation}	
Here, we give an example to illustrate the entanglement swapping of Bell states. Suppose there are two Bell states initially in $|\phi^+\rangle_{12}$ and $|\phi^+\rangle_{34}$, the subscripts $1,2,3,4$ denote the positions of qubits in Bell states. After entanglement swapping between 2 and 3 qubits, then we have 
\begin{equation}
\begin{aligned}
|\phi^+\rangle_{12}|\phi^+\rangle_{34}&=\frac{1}{\sqrt 2} (|00\rangle +|11\rangle)_{12}\otimes\frac{1}{\sqrt 2} (|00\rangle +|11\rangle)_{34}\\
&=\frac{1}{2}(|0000\rangle+|1010\rangle+|0101\rangle +|1111\rangle)_{1324}\\
&=\frac{1}{2}(|\phi^+\rangle|\phi^+\rangle+|\phi^-\rangle|\phi^-\rangle+|\psi^+\rangle|\psi^+\rangle+|\psi^-\rangle|\psi^-\rangle)_{1324}.
\end{aligned}
\end{equation}

Next, let's introduce the prerequisites for the proposed protocol. In our protocol, two classical users, Alice and Bob, are limited to the following operations: (1) \textbf{\emph{measure}}: measure the qubit in basis $\{|0\rangle,|1\rangle\}$ and regenerate one in the same state. (2) \textbf{\emph{reflect}}: reflect the qubit directly. They want to compare their secret information with the help of quantum TP, who has the full quantum capability and may be adversarial. Alice's private binary string is denoted as $M_A=[m^{1}_{A},m^{2}_{A},\dots,m^{n}_{A}]$, while Bob's private binary string is denoted as  $M_B=[m^{1}_{B},m^{2}_{B},\dots,m^{n}_{B}]$. 

\subsection{ Protocol description}
This part gives the detailed steps of the SQPC protocol based on entanglement swapping of Bell states, described in the following procedure. We also provide the flowchart of the proposed SQPC protocol, as shown in Fig. 1.

\begin{figure}[htb]
\centering\includegraphics[width=4.5in]{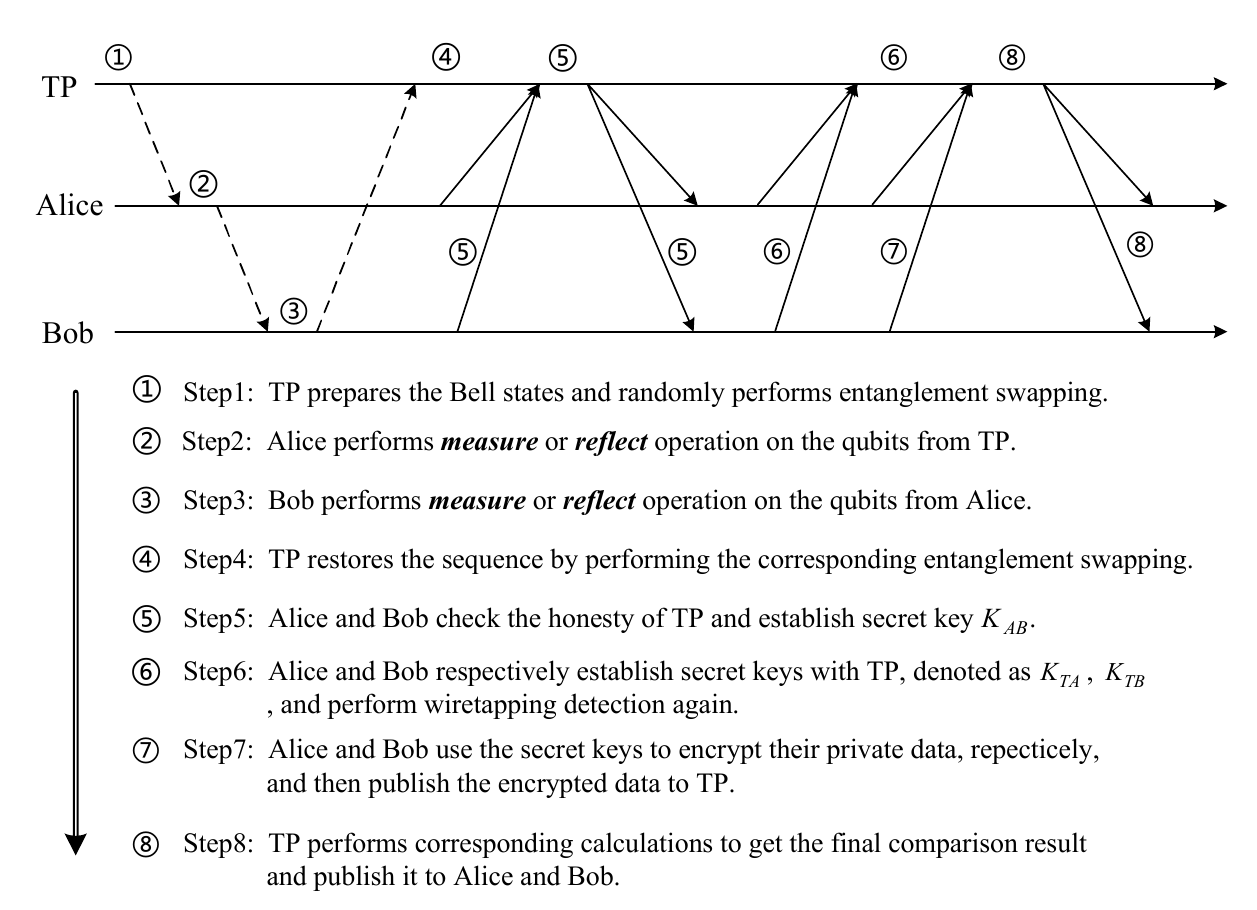}
\caption{The flowchart of our SQPC protocol. In the diagram, the dashed lines indicate the transfer of quantum states, while the solid lines represent the transfer of classical information. }
\label{fig:1}       
\end{figure}
\textbf{Step 1:} TP prepares $4n$ Bell states all in the state of $|\phi^+\rangle=\frac{1}{\sqrt 2}(|00\rangle+|11\rangle)$, and constructs a qubit sequence according to each 2 Bell states as a group, denoted as $S=[G_1,G_2,\dots,G_{2n}]$. Then, TP randomly performs entanglement swapping for each group of 2 Bell states or keeps them unchanged. After that, the sequence $S$ is referred to as $S^*$, and TP sends $S^*$ to Alice.

\textbf {Step 2:} For each received qubit, Alice randomly chooses the \textbf{\emph{measure}} or \textbf{\emph{reflect}} operation. Alice will record the measurement result, when she chooses the \textbf{\emph{measure}} operation. Note that, after Alice manipulates the received qubits, she will send them to Bob.

\textbf {Step 3:} Bob also follows Alice's operation, randomly chooses the \textbf{\emph{measure}} or \textbf{\emph{reflect}} operation. After that,  Bob will send the qubits to TP.

\textbf {Step 4:} After TP receives all the qubits, TP first restores the sequence $S^*$ to its original order, that is, for each group if an entanglement swapping occurred in step 1, then it is resumed here.
\begin{table}[htbp!]
\centering 
\caption{Operations performed by Alice and Bob on the Bell state and the corresponding contribution in steps 5 and 6.  Here, \emph{M} and  \emph{R} mean \textbf{\emph{measure}} and \textbf{\emph{reflect}} operation respectively, \emph{E-C} denotes eavesdropping check. }
\label{tab:1}       
\begin{tabular}{ccccccccccccc}\toprule
Case & \multicolumn{6}{c}{The first qubit of Bell states } 
 & \multicolumn{6}{c}{The second qubit of Bell states }\\ 
  \cmidrule(lr){2-7} \cmidrule(lr){8-13} 
\noalign{\smallskip}
 & \multicolumn{1}{c}{Alice } 
 & \multicolumn{1}{c}{Bob } 
 & \multicolumn{2}{c}{Contribution }
 & \multicolumn{2}{c}{Contribution }  
 & \multicolumn{1}{c}{Alice } 
 & \multicolumn{1}{c}{Bob } 
 & \multicolumn{2}{c}{Contribution }
 & \multicolumn{2}{c}{Contribution }  \\
 & \multicolumn{1}{c}{ } 
 & \multicolumn{1}{c}{ } 
 & \multicolumn{2}{c}{in step5}
 & \multicolumn{2}{c}{in step6 }  
 & \multicolumn{1}{c}{ } 
 & \multicolumn{1}{c}{ } 
 & \multicolumn{2}{c}{in step5 }
 & \multicolumn{2}{c}{in step6 }  \\ 
  \hline 
\noalign{\smallskip}
 1
& \multicolumn{1}{c} {\emph{M}} 
& \multicolumn{1}{c} {\emph{M}} 
& \multicolumn{2}{c} {$K_{AB}$}
& \multicolumn{2}{c} {\emph{E-C}} 
& \multicolumn{1}{c} {\emph{M}} 
& \multicolumn{1}{c} {\emph{M}} 
& \multicolumn{2}{c} {$K_{AB}$}
& \multicolumn{2}{c} {\emph{E-C}} \\
\noalign{\smallskip}
2
& \multicolumn{1}{c} {\emph{M}} 
& \multicolumn{1}{c} {\emph{M}}
& \multicolumn{2}{c} {$K_{AB}$}
& \multicolumn{2}{c} {\emph{E-C}} 
& \multicolumn{1}{c} {\emph{M}}
& \multicolumn{1}{c} {\emph{R}}
& \multicolumn{2}{c} {$-$}
& \multicolumn{2}{c} {$K_{TA}$}   \\
\noalign{\smallskip}
3
& \multicolumn{1}{c} {\emph{M}} 
& \multicolumn{1}{c} {\emph{M}}
& \multicolumn{2}{c} {$K_{AB}$}
& \multicolumn{2}{c} {\emph{E-C}}  
& \multicolumn{1}{c} {\emph{R}}
& \multicolumn{1}{c} {\emph{M}} 
& \multicolumn{2}{c} {$-$}
& \multicolumn{2}{c} {$K_{TB}$}  \\
\noalign{\smallskip}
4
& \multicolumn{1}{c} {\emph{M}} 
& \multicolumn{1}{c} {\emph{M}}
& \multicolumn{2}{c} {$K_{AB}$}
& \multicolumn{2}{c} {\emph{E-C}}  
& \multicolumn{1}{c} {\emph{R}}
& \multicolumn{1}{c} {\emph{R}}
& \multicolumn{2}{c} {$-$}
& \multicolumn{2}{c} {$-$}    \\
\noalign{\smallskip}
5
& \multicolumn{1}{c} {\emph{M}} 
& \multicolumn{1}{c} {\emph{R}} 
& \multicolumn{2}{c} {$-$}
& \multicolumn{2}{c} {$K_{TA}$}  
& \multicolumn{1}{c} {\emph{M}}
& \multicolumn{1}{c} {\emph{M}}
& \multicolumn{2}{c} {$K_{AB}$}
& \multicolumn{2}{c} {\emph{E-C}}    \\
\noalign{\smallskip}
6
& \multicolumn{1}{c} {\emph{M}} 
& \multicolumn{1}{c} {\emph{R}} 
& \multicolumn{2}{c} {$-$}
& \multicolumn{2}{c} {$K_{TA}$} 
& \multicolumn{1}{c} {\emph{M}}
& \multicolumn{1}{c} {\emph{R}}
& \multicolumn{2}{c} {$-$}
& \multicolumn{2}{c} {$K_{TA}$}   \\
\noalign{\smallskip}
7
& \multicolumn{1}{c} {\emph{M}} 
& \multicolumn{1}{c} {\emph{R}}
& \multicolumn{2}{c} {$-$}
& \multicolumn{2}{c} {$K_{TA}$} 
& \multicolumn{1}{c} {\emph{R}}
& \multicolumn{1}{c} {\emph{M}}
& \multicolumn{2}{c} {$-$}
& \multicolumn{2}{c} {$K_{TB}$}   \\
\noalign{\smallskip}
8
& \multicolumn{1}{c} {\emph{M}} 
& \multicolumn{1}{c} {\emph{R}} 
& \multicolumn{2}{c} {$-$}
& \multicolumn{2}{c} {$K_{TA}$} 
& \multicolumn{1}{c} {\emph{R}}
& \multicolumn{1}{c} {\emph{R}} 
& \multicolumn{2}{c} {$-$}
& \multicolumn{2}{c} {$-$}  \\
\noalign{\smallskip}
9
& \multicolumn{1}{c} {\emph{R}} 
& \multicolumn{1}{c} {\emph{M}} 
& \multicolumn{2}{c} {$-$}
& \multicolumn{2}{c} {$K_{TB}$}
& \multicolumn{1}{c} {\emph{M}}
& \multicolumn{1}{c} {\emph{M}}
& \multicolumn{2}{c} {$K_{AB}$}
& \multicolumn{2}{c} {\emph{E-C}}  \\
\noalign{\smallskip}
10
& \multicolumn{1}{c} {\emph{R}} 
& \multicolumn{1}{c} {\emph{M}}
& \multicolumn{2}{c} {$-$}
& \multicolumn{2}{c} {$K_{TB}$}
& \multicolumn{1}{c} {\emph{M}}
& \multicolumn{1}{c} {\emph{R}} 
& \multicolumn{2}{c} {$-$}
& \multicolumn{2}{c} {$K_{TA}$} \\
\noalign{\smallskip}
11
& \multicolumn{1}{c} {\emph{R}} 
& \multicolumn{1}{c} {\emph{M}} 
& \multicolumn{2}{c} {$-$}
& \multicolumn{2}{c} {$K_{TB}$}
& \multicolumn{1}{c} {\emph{R}}
& \multicolumn{1}{c} {\emph{M}}
& \multicolumn{2}{c} {$-$}
& \multicolumn{2}{c} {$K_{TB}$}  \\
\noalign{\smallskip}
12
& \multicolumn{1}{c} {\emph{R}} 
& \multicolumn{1}{c} {\emph{M}} 
& \multicolumn{2}{c} {$-$}
& \multicolumn{2}{c} {$K_{TB}$} 
& \multicolumn{1}{c} {\emph{R}}
& \multicolumn{1}{c} {\emph{R}}
& \multicolumn{2}{c} {$-$}
& \multicolumn{2}{c} {$-$}   \\
\noalign{\smallskip}
13
& \multicolumn{1}{c} {\emph{R}} 
& \multicolumn{1}{c} {\emph{R}} 
& \multicolumn{2}{c} {$-$}
& \multicolumn{2}{c} {$-$} 
& \multicolumn{1}{c} {\emph{M}}
& \multicolumn{1}{c} {\emph{M}}
& \multicolumn{2}{c} {$K_{AB}$}
& \multicolumn{2}{c} {\emph{E-C}} \\
\noalign{\smallskip}
14
& \multicolumn{1}{c} {\emph{R}} 
& \multicolumn{1}{c} {\emph{R}} 
& \multicolumn{2}{c} {$-$}
& \multicolumn{2}{c} {$-$}
& \multicolumn{1}{c} {\emph{M}}
& \multicolumn{1}{c} {\emph{R}}
& \multicolumn{2}{c} {$-$}
& \multicolumn{2}{c} {$K_{TA}$}  \\
\noalign{\smallskip}
15
& \multicolumn{1}{c} {\emph{R}} 
& \multicolumn{1}{c} {\emph{R}} 
& \multicolumn{2}{c} {$-$}
& \multicolumn{2}{c} {$-$}
& \multicolumn{1}{c} {\emph{R}}
& \multicolumn{1}{c} {\emph{M}}
& \multicolumn{2}{c} {$-$}
& \multicolumn{2}{c} {$K_{TB}$}  \\
\noalign{\smallskip}
16
& \multicolumn{1}{c} {\emph{R}} 
& \multicolumn{1}{c} {\emph{R}}
& \multicolumn{2}{c} {\emph{E-C}}
& \multicolumn{2}{c} {\emph{E-C}} 
& \multicolumn{1}{c} {\emph{R}}
& \multicolumn{1}{c} {\emph{R}}
& \multicolumn{2}{c} {\emph{E-C}}
& \multicolumn{2}{c} {\emph{E-C}} \\
\bottomrule
\end{tabular} 
\end{table}

\textbf {Step 5:} Alice and Bob will discuss eavesdropping and TP's honesty in this step, while also establishing the secret key. In more detail, Alice and Bob first randomly select half of the Bell state groups, and then let TP measure the qubits in Bell basis and publish the measurement results. Finally, Alice and Bob publish their operations on these selected qubits and discuss the results. The following cases need to be focused on. 
\begin{itemize}
\item[(1)] Alice and Bob both choose the \textbf{\emph{reflect}} operation on the qubits of Bell states, and in this case TP should always publish $|\phi^+\rangle$. This case is used for checking the honesty of TP and eavesdropping. If error rate surpasses the threshold, the protocol ends.

\item[(2)] Alice and Bob choose the \textbf{\emph{measure}} operation on the same qubits of Bell states, and TP publishes the result as one of four Bell states (i.e., $|\phi^+\rangle$,$|\phi^-\rangle$,$|\psi^+\rangle$,$|\psi^-\rangle$, because the qubits Alice and Bob measured may be after the entanglement swapping). In this case, Alice and Bob's measurement result should be the same, while TP knows nothing about their measurements because TP performs Bell-based measurements, not $Z$-based measurements. Thus, Alice and Bob can establish a raw secret key sequence denoted as $K_{AB}=[k^{1}_{AB},k^{2}_{AB},\dots,k^{n}_{AB}]$.
\end{itemize}
Table 1 lists the cases where Alice and Bob choose different operations for the Bell state, and gives the corresponding contributions of different operations in steps 5 and 6.

\textbf {Step 6:} For the remaining groups, Alice and Bob first tell TP their operations on these qubits, and then, TP measures these qubits in $Z$ basis or Bell basis according to the published information. As a result, the following scenarios will happen:
\begin{itemize}
\item {1)} For the two qubits belonging to a Bell state, Alice and Bob both perform the \textbf{\emph{reflect}} operation, TP measures these qubits in Bell basis. If there is no Eve online, TP's measurement result must be $|\phi^+\rangle$. If the error rate exceeds the threshold, the protocol terminates and restarts.

\item {2)} For the qubits that Alice and Bob choose different operations, TP measures the qubits in $Z$ basis. In this way, TP can establish secret key with Alice and Bob, respectively. More specifically, if Alice chooses \textbf{\emph{measure}} operation and Bob chooses the \textbf{\emph{reflect}} operation, then Alice and TP can establish secret key based on their measurement results. Bob and TP can share the secret key with the similar way. We use $K_{TA}=[k^{1}_{TA},k^{2}_{TA},\dots,k^{n}_{TA}]$ ($K_{TB}=[k^{1}_{TB},k^{2}_{TB},\dots,k^{n}_{TB}]$) to represent the secret key between Alice and TP (Bob and TP).

\item {3)} For the qubits that Alice and Bob both choose the \textbf{\emph{measure}} operation, TP will also measure the qubits in $Z$ basis, and all of three users' measurement results should be the same.  If the error rate exceeds the threshold, the protocol terminates and restarts.

\end{itemize}

\textbf {Step 7:} All three users have established secret key with each other. Then, Alice and Bob will use the secret keys obtain from steps 5 and 6 to encrypt their private data, respectively. In detail, Alice and Bob calculate $Q^{j}_{A}=k^{j}_{AB}\oplus k^{j}_{TA}\oplus m^{j}_{A}$ and $Q^{j}_{B}=k^{j}_{AB}\oplus k^{j}_{TB}\oplus m^{j}_{B}$, respectively, where $\oplus$ is the modulo 2 summation, and $j\in\{1,2,\dots,n\}$. Afterwards, Alice and Bob send $Q_A=[Q^{1}_{A},Q^{2}_{A},\dots,Q^{n}_{A}]$ and $Q_B=[Q^{1}_{B},Q^{2}_{B},\dots,Q^{n}_{B}]$ to TP.

\textbf {Step 8:} TP calculates $R^{j}=Q^{j}_{A}\oplus Q^{j}_{B}\oplus k^{j}_{TA}\oplus k^{j}_{TB}$. If $R^j=0$ for $j=1,2,\dots,n$, she will conclude that the secrets of Alice and Bob are equal. Otherwise, their secrets are not equal. Finally, TP announces the comparison result to Alice and Bob.

\subsection {Correctness of protocol}
In our protocol, Alice and Bob can establish the secure key sequence $K_{AB}=[k^{1}_{AB},k^{2}_{AB},\dots,k^{n}_{AB}]$. Then, TP can establish the secure key sequence $K_{TA}=[k^{1}_{TA},k^{2}_{TA},\dots,k^{n}_{TA}]$ ($K_{TB}=[k^{1}_{TB},k^{2}_{TB},\dots,k^{n}_{TB}]$) with Alice (Bob). Consequently, Alice uses $K^{j}_{AB}$ and $K^{j}_{TA}$ to encrypt her secret information $m^j_A$ as 
\begin{equation}
Q^{j}_{A}=k^{j}_{AB}\oplus k^{j}_{TA}\oplus m^{j}_{A}. 
\end{equation}
Bob also uses $K^{j}_{AB}$ and $K^{j}_{TB}$ to encrypt her secret information $m^j_B$ as 

\begin{equation}
Q^{j}_{B}=k^{j}_{AB}\oplus k^{j}_{TB}\oplus m^{j}_{B}. 
\end{equation}
Finally, TP calculates $R^{j}=Q^{j}_{A}\oplus Q^{j}_{B}\oplus k^{j}_{TA}\oplus k^{j}_{TB}$. It is easy to obtain that 
\begin{equation}
\begin{aligned}
R^{j}&=Q^{j}_{A}\oplus Q^{j}_{B}\oplus k^{j}_{TA}\oplus k^{j}_{TB}\\
&=k^{j}_{AB}\oplus k^{j}_{TA}\oplus m^{j}_{A}\oplus k^{j}_{AB}\oplus k^{j}_{TB}\oplus m^{j}_{B}\oplus k^{j}_{TA}\oplus k^{j}_{TB}\\
&=m^{j}_{A}\oplus m^{j}_{B}.
\end{aligned}
\end{equation}
If $R^j=0$ for $j=1,2,\dots,n$, she will infer that the secrets of Alice and Bob are equal. Otherwise, their secrets are not equal. The results show that the proposed protocol can guarantee the correctness of the output.

\section{Security analysis}

In the above section, we gave the specific steps of the SQPC protocol and realized the privacy comparison. Let us now investigate the security of the proposed protocol. In more detail, we will show that the attacks from the outside eavesdroppers (i.e., the external attack) and the attacks from the adversarial users and TP (i.e., the internal attack) are invalid.

\subsection{External attack}
In the following, we show that Eve will be captured regardless of the attack she uses, such as the intercept-resend attack,  measure-resend attack, collective attack, and Double-CNOT attack. While not all categories of attacks are exhaustively covered, the analysis includes most of typical attacks usually considered in SQPC protocols.

\subsubsection{The intercept-resend attack}
Suppose an external eavesdropper, Eve, uses the intercept-resend attack to obtain some useful information. There are two kind of intercept-resend attacks according to the process of the proposed protocol. We will do the detailed analysis for them in the follows.

First, Eve may intercept all the qubits sent from TP to Alice and return them to TP directly. Then, she generates fake qubits in basis $ \{|0\rangle, |1\rangle \}$, and sends these fake qubits to Alice. After Alice and Bob performing the operations on these qubits, Eve catches these qubits from Bob and measures them in basis $ \{|0\rangle, |1\rangle \}$ to obtain some useful information. On the one hand, if Alice and Bob both choose the \textbf{\emph{reflect}} operation on the received qubits, Eve's attack will not be detected. On the other hand, if Alice and Bob both choose the \textbf{\emph{measure}} operation on the qubits, and TP also measures them in $Z$ basis, then Eve's attack will be discovered with non-zero probability in step 6. This is because the measurement results of fake qubits prepared by Eve cannot be perfectly consistent with the results of TP. Therefore, TP, Alice and Bob cannot obtain the same results, indicating that Eve's attack fail. 

Second, Eve may intercept all the qubits sent from Alice to Bob and return them to TP directly. Then, she generates fake qubits in basis $ \{|0\rangle, |1\rangle \}$, and sends these fake qubits to Bob. After Bob performing the operations on these qubits, Eve catches these qubits from Bob and measures them in basis $ \{|0\rangle, |1\rangle \}$ to obtain some useful information. Despite this, Eve's attack was still unable to evade eavesdropping detection and was eventually discovered by TP. If Alice and Bob both choose the \textbf{\emph{measure}} operation on the qubits, and TP also measures them in $Z$ basis, then the fake qubits prepared by Eve will inevitably cause measurement results between Alice, Bob, and TP to be inconsistent. Therefore, Eve's attack will inevitably introduce errors and be detected by the eavesdropping detection mechanism of the protocol.

\subsubsection{The measure-resend attack}
In our protocol, Eve may launch the measure-resend attack to steal some useful information. The measure-resend attack here is that Eve may perform measurements in different bases to obtain some useful information. We focus on analyzing the two typical cases where Eve uses the $Z$ basis and Bell basis for measurements. The analysis using other measurement bases is similar.

Assume Eve first uses $Z$ basis to measure the intercepted qubits from TP, Alice, or Bob. Afterwards, she sends the measured qubits to the next recipient. However, since Eve does not know which operations Alice and Bob have employed, she can only avoid detection if either Alice or Bob decides to perform \textbf{\emph{measure}} operation. When Alice and Bob both choose the \textbf{\emph{reflect}} operation on two qubits of 
a Bell state, then Eve's attack will be exposed in steps 5 and 6, as her measurement has destroyed the entanglement of the Bell state.

Next, suppose Eve uses Bell basis to measure the intercepted qubits. However, through this kind of attack, Eve will also inevitably be detected in steps 5 and 6, as her measurement may destroy the entanglement swapping of Bell states. Specifically, for the transmitted quantum sequence, TP randomly performs entanglement swapping on two Bell states in each group or keeps them unchanged. This means that Eve does not know whether the Bell states in each transmitted group have undergone entanglement swapping or not. If Eve rashly selects particles for Bell basis measurement, her measurement may disrupt the entanglement swapping between the Bell states and consequently be detected by the TP. Therefore, adopting the measure-resend attack, Eve will unavoidably be found by TP.

\subsubsection{The collective attack}
The collective attack is a type of quantum attack where the attacker repetitively performs the same attack at each iteration, while keeping the quantum information in their ancillary for later measurements. It can often lead to more successful eavesdropping than individual attacks on the transmitted qubits \cite{34,35}. We will show our protocol can withstand the collective attack in the following.

Our protocol relies on three one-way quantum channels to realize the transmission of the whole qubits, so the collective attack can be modeled as three unitaries $U_1$, $U_2$ and $U_3$ \cite{36,37}. Here, $U_1$, $U_2$, and $U_3$ are the attack operators applied to the qubits sent between TP-Alice, Alice-Bob, and Bob-TP, respectively (see Fig. 2). Note that $U_1$, $U_2$, and $U_3$ share a common probe space with initial state $|0\rangle_E$. The detailed analysis is shown below.
\begin{figure}[htb]
\centering\includegraphics[width=3.5in]{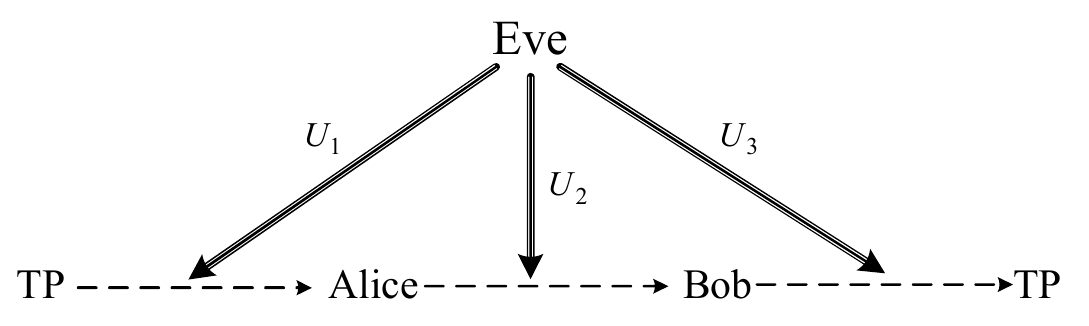}
\caption{A schematic of Eve's collective attack. Eve will launch $U_1$, $U_2$, and $U_3$ on the qubits from TP to Alice, Alice to Bob, and Bob to TP, respectively, to obtain some helpful information.}
\label{fig:2}       
\end{figure}

\begin{theorem}
Assume that Eve launch attack $(U_1,U_2,U_3)$ on the target qubit and her probe $|0\rangle_E$. For this attack inducing no error, the final state of Eve's probe should be independent of the target qubit. That is, Eve obtains nothing about the secret information.
\end{theorem}

\noindent \textbf{Proof.} In our protocol, the basic information carrier for transmission is the Bell state (for Eve, she cannot distinguish whether entanglement swapping has occurred between the Bell states).  Therefore, in the following analysis, we take the Bell state as the basic unit to analyze the effect of $U_1$, $U_2$ and $U_3$.  We specifically focus on the cases where Alice and Bob both choose \textbf{\emph{measure}} or \textbf{\emph{reflect}} operation on the Bell state, as the protocol sets up corresponding detection mechanisms in these cases to detect Eve' attacks. 

We begin by analyzing the scenario where both Alice and Bob choose to perform \textbf{\emph{measure}} operation and Eve launches attacks without introducing any errors.

1) Eve first performs $U_1$ on the Bell state from TP to Alice.
The effect of $U_1$ on the qubits $|0\rangle$ and $|1\rangle$ is:
\begin{equation}
\begin{aligned}
U_1|0,0\rangle_{TE} =\alpha_0|0,E_0\rangle + \alpha_1|1,E_1\rangle,\\
U_1|1,0\rangle_{TE} =\alpha_2|0,E_2\rangle + \alpha_3|1,E_3\rangle,
\end{aligned}
\end{equation}
where $T$ and $E$ represent the target qubit and Eve's probe, respectively. $|E_0\rangle$, $|E_1\rangle$, $|E_2\rangle$, $|E_3\rangle$ are the pure states determined by $U_1$, and $\alpha_0^2+\alpha_1^2=1$, $\alpha_2^2+\alpha_3^2=1$. Thus, after Eve performing $U_1$ on the two qubits of Bell state $|\phi^+\rangle$ and her probe, respectively, we have
\begin{equation}
\frac{1}{\sqrt 2}\left[
\begin{aligned}
|00\rangle (\alpha_0\alpha_0|E_0E_0\rangle+\alpha_2\alpha_2|E_2E_2\rangle)+|10\rangle (\alpha_1\alpha_0|E_1E_0\rangle+\alpha_3\alpha_2|E_3E_2\rangle)\\
+|01\rangle (\alpha_0\alpha_1|E_0E_1\rangle+\alpha_2\alpha_3|E_2E_3\rangle)+|11\rangle (\alpha_1\alpha_1|E_1E_1\rangle+\alpha_3\alpha_3|E_3E_3\rangle)
\end{aligned}
\right].
\end{equation}
After rearranging (7), it can be expressed as 
\begin{equation}
\frac{1}{\sqrt 2}\sum_{i,j=0,1} |i,j\rangle (\alpha_i\alpha_j|E_iE_j\rangle+\alpha_{i+2}\alpha_{j+2}|E_{i+2}E_{j+2}\rangle).
\end{equation}
From (8), we can see that if Alice's measurement results are $|i\rangle$ and $|j\rangle$, then Eve's probe will be $(|E_iE_j\rangle+|E_{i+2}E_{j+2}\rangle)$. After Alice performing \textbf{\emph{measure}} operation, she will send new qubits based on her measurement results to Bob.

2) Eve then launches $U_2$ on the qubits $|i,j\rangle$ from Alice to Bob. The effect $U_2$ on the qubit $|i\rangle$ (or $|j\rangle$) and the corresponding probe is:
\begin{equation}
\begin{aligned}
&U_2|i,E_i\rangle = \sum_{k=0,1}\alpha_{k,i}|k,E_{k,i}\rangle,\\
&U_2|i,E_{i+2}\rangle =\sum_{k=0,1}\alpha_{k,i+2}|k,E_{k,i+2}\rangle,\\
&U_2|j,E_j\rangle =\sum_{l=0,1}\alpha_{l,j}|l,E_{l,j}\rangle,\\
&U_2|j,E_{j+2}\rangle =\sum_{l=0,1}\alpha_{l,j+2}|l,E_{l,j+2}\rangle,
\end{aligned}
\end{equation}
where $\sum_{k=0,1}\alpha^2_{k,i}=1$, $\sum_{k=0,1}\alpha^2_{k,i+2}=1$, $\sum_{l=0,1}\alpha^2_{l,j}=1$, and $\sum_{l=0,1}\alpha^2_{l,j+2}=1$. $|E_{k,i}\rangle$, $|E_{k,i+2}\rangle$, $|E_{l,j}\rangle$, $|E_{l,j+2}\rangle$ are the pure states determined by $U_2$. Thus, after Eve performing $U_2$ on each of the $|i,j\rangle$ and her probe, we have (ignoring the normalization coefficient)
\begin{equation}
\begin{aligned}
\sum_{k,l=0,1} |k,l\rangle (\alpha_{k,i}\alpha_{l,j}|E_{k,i}E_{l,j}\rangle+\alpha_{k,i+2}\alpha_{l,j+2}|E_{k,i+2}E_{l,j+2}\rangle).
\end{aligned}
\end{equation}
From (10), we can see that if Bob's measurement results are $|k\rangle$ and $|l\rangle$, then Eve's probe will be $(|E_{k,i}E_{l,j}\rangle+|E_{k,i+2}E_{l,j+2}\rangle)$. After Bob performing \textbf{\emph{measure}} operation, he will send his measured qubits to TP.

3) Eve finally launches $U_3$ on the qubits $|k,l\rangle$ from Bob to TP. The effect $U_3$ on the qubit $|k\rangle$ (or $|l\rangle$) and Eve's probe is:
\begin{equation}
\begin{aligned}
&U_3|k,E_{k,i}\rangle =\sum_{g=0,1}\alpha_{g,k,i}|g,E_{g,k,i}\rangle,\\
&U_3|k,E_{k,i+2}\rangle =\sum_{g=0,1}\alpha_{g,k,i+2}|g,k,E_{k,i+2}\rangle,\\
&U_3|l,E_{l,j}\rangle =\sum_{h=0,1}\alpha_{h,l,j}|h,E_{h,l,j}\rangle,\\
&U_3|l,E_{l,j+2}\rangle =\sum_{h=0,1}\alpha_{h,l,j+2}|h,E_{h,l,j+2}\rangle,
\end{aligned}
\end{equation}
where $\sum_{g=0,1}\alpha^2_{g,k,i}=1$, $\sum_{g=0,1}\alpha^2_{g,k,i+2}=1$, $\sum_{h=0,1}\alpha^2_{h,l,j}=1$, and $\sum_{h=0,1}\alpha^2_{h,l,j+2}=1$. $|E_{g,k,i}\rangle$, $|E_{g,k,i+2}\rangle$, $|E_{h,l,j}\rangle$, $|E_{h,l,j+2}\rangle$ are the pure states determined by $U_3$.
Thus, after Eve performing $U_3$ on each of the $|k,l\rangle$ and her probe, we have 
\begin{equation}
\begin{aligned}
\sum_{g,h=0,1} |g,h\rangle (\alpha_{g,k,i}\alpha_{h,l,j}|E_{k,i}E_{l,j}\rangle+\alpha_{g,k,i+2}\alpha_{h,l,j+2}|E_{g,k,i+2}E_{h,l,j+2}\rangle).
\end{aligned}
\end{equation}
After TP received the qubits from Bob, she will measure these qubits in $Z$ basis according to step 6, when Alice and Bob both choose the \textbf{\emph{measure}} operation. In this case, if Eve's attack introduces no errors, the measurement results of the three users should be consistent, so we have
\begin{equation}
\begin{aligned}
|i\rangle=|k\rangle=|g\rangle, \quad |j\rangle=|l\rangle=|h\rangle, \quad i.e.,\quad  g=k=i,\quad  h=l=j.
\end{aligned}
\end{equation} 

Furthermore, as the Bell state sent by TP is $|\phi^+\rangle$, when Alice and Bob choose to perform \textbf{\emph{measure}} operation on the two qubits of $|\phi^+\rangle$, their measurement outcomes for the two qubits should be the same. Thus, combined with (13), we can get
\begin{equation}
  g=k=i=h=l=j.
\end{equation} 
From the above analysis, it can be concluded that if Eve wants to avoid introducing any errors, equation (14) must be satisfied.  This corresponds to the operations $(U_1,U_2,U_3)$ performed by Eve, which must meet the following conditions. Specifically, for $U_1$, the result of its action on the target qubit should satisfy (i.e., (6) should satisfy): 
\begin{equation}
U_1|0,0\rangle_{TE} =|0,E_0\rangle,  \quad U_1|1,0\rangle_{TE} =|1,E_3\rangle, \quad \alpha_0=\alpha_3=1,\quad \alpha_1=\alpha_2=0.
\end{equation}
Following (15),  it leads to the transformation of (8) into 
\begin{equation}
\frac{1}{\sqrt 2}( |0,0\rangle|E_0E_0\rangle+ |1,1\rangle|E_3E_3\rangle),
\end{equation}
to ensure that Alice's measurement results on the two qubits of $|\phi^+\rangle$ remain consistent. 

Then for $U_2$, it must meet the similar requirements of $U_1$, that is, the effect of $U_2$ cannot alter the states of the qubits from Alice to Bob. Suppose Alice's measurement result is $|0,0\rangle$, then the corresponding probe of Eve will be $|E_0E_0\rangle$. In order to avoid introducing errors in $U_2$ attack, in this case Bob's observed result must be $|0,0\rangle$. The situation is similar when Alice obtains $|1,1\rangle$ and Eve's probe becomes $|E_3E_3\rangle$, where Bob's observed result should be $|1,1\rangle$. Therefore, the effect of $U_2$ should satisfy:
\begin{equation}
\begin{aligned}
U_2|0\rangle|E_{0}\rangle=|0\rangle|E_{0,0}\rangle, \quad U_2|1\rangle|E_{3}\rangle=|1\rangle|E_{1,3}\rangle, \quad   \alpha_{0,0}=\alpha_{1,3}=1,\quad \alpha_{1,0}=\alpha_{0,3}=0,
\end{aligned}
\end{equation}
so after the attack of $U_2$, the system will be  
\begin{equation}
\left\{
\begin{aligned}
&|0,0\rangle|E_{0,0}E_{0,0}\rangle,\quad  \text{if Alice observes }|0,0\rangle, \\
&|1,1\rangle|E_{1,3}E_{1,3}\rangle, \quad  \text{if Alice observes }|1,1\rangle .
\end{aligned}
\right.
\end{equation}
Only then can it be guaranteed that the result observed by Bob is consistent with Alice's.

Finally, with regards to $U_3$, its effect should also not change the state of the qubit sent by Bob to TP. Assuming the measurement results of both Alice and Bob are $|0,0\rangle$, and then Eve's probe becomes $|E_{0,0}E_{0,0}\rangle$.  In this case, $U_3$ goes undetected only if 
TP's measurement result is $|0,0\rangle$. When Alice and Bob obtain $|1,1\rangle$, the situation is similar, and the measurement result of TP should be $|11\rangle$ in this case. Thus, $U_3$ should meet the following conditions:
\begin{equation}
U_3|0\rangle|E_{0,0}\rangle=|0\rangle|E_{0,0,0}\rangle, \quad U_3 |1\rangle|E_{1,3}\rangle=|1\rangle|E_{1,1,3}\rangle, \quad \alpha_{1,0,0}=\alpha_{0,1,3}=0,\quad \alpha_{0,0,0}=\alpha_{1,1,3}=1.
\end{equation}
Following (19), after undergoing $U_3$ the system will become 
\begin{equation}
\left\{
\begin{aligned}
&|0,0\rangle|E_{0,0,0}E_{0,0,0}\rangle,\quad  \text{if Alice and Bob observe }|0,0\rangle, \\
&|1,1\rangle|E_{1,1,3}E_{1,1,3}\rangle, \quad  \text{if Alice and Bob observe }|1,1\rangle ,
\end{aligned}
\right.
\end{equation}
so that the measurement result of TP is consistent with that of Alice and Bob.

Next, we analyze the scenario where both Alice and Bob choose to perform \textbf{\emph{reflect}} and Eve launches attacks without introducing any errors.

i) Eve first performs $U_1$ on the Bell state from TP to Alice. As analyzed above, after the Bell state is attacked by $U_1$, if no error is introduced, the state will become (16). After that, Alice will reflect the received qubits to Bob.

ii) Eve then launches $U_2$ on the qubits from Alice to Bob. According to (17), after performing $U_2$, the system's state will be 
\begin{equation}
\begin{aligned}
\frac{1}{\sqrt 2}(|00\rangle|E_{0,0}E_{0,0}\rangle+|11\rangle|E_{1,3}E_{1,3}\rangle).
\end{aligned}
\end{equation}
Bob also chooses the \textbf{\emph{reflect}} operation, so he returns the received qubits directly to TP.

iii) Eve finally launches $U_3$ on the qubits from Bob to TP. Following (19), after performing $U_3$, the system's state becomes
\begin{equation}
\frac{1}{\sqrt 2}(|00\rangle|E_{0,0,0}E_{0,0,0}\rangle+|11\rangle|E_{1,1,3}E_{1,1,3}\rangle).
\end{equation}
After TP received the qubits from Bob, she will perform the measurement on the Bell basis when Alice and Bob both choose \textbf{\emph{reflect}} operation. In this case, if Eve wants to be undetected in steps 5 and 6, TP's measurement result should always be $|\phi^+\rangle$. Therefore, we have
\begin{equation}
|E_{0,0,0}\rangle=|E_{1,1,3}\rangle.
\end{equation}
So that (22) can be written as 
\begin{equation}
\frac{1}{\sqrt 2}(|00\rangle+|11\rangle)|E_{0,0,0}E_{0,0,0}\rangle=|\phi^+\rangle|E_{0,0,0}E_{0,0,0}\rangle.
\end{equation}
That is to say, in the scenario where both Alice and Bob choose the \textbf{\emph{reflect}} operation, Eve's probe remains in the same state $|E_{0,0,0}\rangle$ after Eve applying $(U_1,U_2,U_3)$ attacks on the target qubit.

Recall the scenario where both Alice and Bob choose the \textbf{\emph{measure}} operation. After Eve applying $(U_1,U_2,U_3)$ attacks on the target qubit, the system's state will become to (20). Then, apply (23) into (20), we can obtain
\begin{equation}
\left\{
\begin{aligned}
&|0,0\rangle|E_{0,0,0}E_{0,0,0}\rangle \\
&|1,1\rangle|E_{1,1,3}E_{1,1,3}\rangle=|1,1\rangle|E_{0,0,0}E_{0,0,0}\rangle,
\end{aligned}
\right.
\end{equation}
i.e., in the scenario where both Alice and Bob choose the \textbf{\emph{measure}} operation, Eve's probe remains in the same state $|E_{0,0,0}\rangle$ no matter what state the target qubit is in. This also means that Eve's probe is independent of the target qubit.

To combine the two scenarios, it is clear that if Eve intends to avoid introducing errors, Eve can only get the same result from her probe (i.e., her probe must always remain independent of the target qubits), regardless of the operation chosen by Alice and Bob. Therefore, our protocol can withstand the collective attack.

\subsubsection{The Double-CNOT attack}
As our protocol relies on multiple quantum channels, it is susceptible to the Double-CNOT attack. 
Assuming that Alice and Bob play identical roles for Eve, then Eve may perform the following steps to eavesdrop on Alice's secret information: 

1) Eve firstly intercepts the qubit from TP to Alice. Then, she performs the first attack $U_{CNOT}=|00\rangle \langle00|+|01\rangle \langle01|+|10\rangle \langle10|+|11\rangle \langle11|$, where the intercepted qubit is the control qubit and her ancillary particle $\left|0\right\rangle_E$ is the target qubit.

 2) Eve intercepts the qubit from Alice to Bob and executes the second $U_{CNOT}$ on the intercepted qubit and her ancillary particle again. Finally, Eve measures her ancillary particle to obtain some useful information.

However, Eve's Double-CNOT attack cannot succeed in our protocol. In the following, we analyze the effect of Double-CNOT attack on the qubit of the Bell state. Concretely speaking, for the qubit $|0\rangle$ and $|1\rangle$, after the first CNOT attack, the composite system will be:
\begin{equation}
\begin{aligned}
U_{CNOT} \big(|0\rangle_T |0\rangle_E \big) =|00\rangle_{TE},\quad U_{CNOT} \big(|1\rangle_T |0\rangle_E \big) =|11\rangle_{TE},
\end{aligned}
\end{equation}
where the subscripts $T$ and $E$ represent the target qubit and Eve's ancillary particle, respectively.  Let us analyze the Bell state situation. 
Here, we assume that Eve attacks the first qubit of the Bell state. After the first CNOT attack, the composite system will be:
\begin{equation}
\begin{aligned}
&U_{CNOT} \left[\frac{1}{\sqrt 2}(|00\rangle_{TE}|0\rangle+|10\rangle_{TE}|1\rangle) \right]=\frac{1}{\sqrt 2}(|00\rangle_{TE}|0\rangle+|11\rangle_{TE}|1\rangle),
\end{aligned}
\end{equation}
where $U_{CNOT}$ is performed on the qubits $T$ and $E$. 
Then, Alice performs the \textbf{\emph{measure}} operation or the \textbf{\emph{reflect}} operation on the received qubits.

If Alice chooses the \textbf{\emph{measure}} operation and obtains the measurement result $|0\rangle$ ($|1\rangle$), then after Eve's the second CNOT attack, the system becomes
\begin{equation}
\begin{aligned}
&U_{CNOT} |00\rangle_{TE}|0\rangle= |00\rangle_{TE}|0\rangle,\quad  \text{if Alice obtains }|0\rangle, \\
&U_{CNOT} |11\rangle_{TE}|1\rangle= |10\rangle_{TE}|1\rangle, \quad  \text{if Alice obtains }|1\rangle .
\end{aligned}
\end{equation}
If Alice chooses the \textbf{\emph{reflect}} operation, then after Eve's the second CNOT attack, the system becomes
\begin{equation}
\begin{aligned}
&U_{CNOT} \left[\frac{1}{\sqrt 2}(|00\rangle_{TE}|0\rangle+|11\rangle_{TE}|1\rangle)\right]=\frac{1}{\sqrt 2}(|00\rangle_{TE}|0\rangle+|10\rangle_{TE}|1\rangle)\\
&\quad =\frac{1}{\sqrt 2}(|00\rangle+|11\rangle)|0\rangle_E.
\end{aligned}
\end{equation}
From (28-29), it easy to obtain that Eve's ancillary particle is always in the state of $|0\rangle$. That is to say, Eve cannot obtain some useful information from her ancillary particle. Therefore, Eve's Double-CNOT attack for our protocol is invalid.


\subsection{Internal attack}
The security of SQPC protocol may face a greater threat from adversarial users and TP compared to external attackers.  Specifically, two scenarios merit consideration: one is that an adversarial user attempts to eavesdrop on the secret from another, the other is that TP wants to obtain the secret from the classical users.

\subsubsection{The attack from one adversarial user}
In the proposed protocol, Alice and Bob play the same role, either of whom may be adversarial. Without loss of generality, we suppose Alice is adversarial and she wants to obtain Bob's secret information. Because Alice already has the secret key $K_{AB}$, she can steal Bob's private data only by obtaining the secret key $K_{TB}$.

However, in our protocol, Bob's operations on the received qubits are randomized and completely independent of Alice's knowledge. Therefore, if Alice aims to obtain the secret information between Bob and TP, she needs to perform attacks on the qubits sent by Bob to TP, similar to an external eavesdropper like Eve. Alice may employ attacks such as the intercept-resend attack and the collective attack to obtain useful information. In this scenario, Alice essentially acts as an external eavesdropper, and according to the above analysis of external attacks, Bob and TP can detect Alice's attacks with a non-zero probability.  It should be noted that although the basic information carrier of the protocol is $|\phi^+\rangle$, Alice cannot infer Bob's measurement result from her own measurement result (e.g., in cases 7 and 10 of Table 1) because she does not know whether TP has performed entanglement swapping on the Bell state. If entanglement swapping is performed, there is no deterministic correspondence between measurement results of Alice and Bob.

Furthermore, Alice can obtain the secure key sequences $K_{AB}$ and $K_{TA}$, and the comparison result.  However, it is still helpless for her to get Bob's secret. Because Bob's secret is encrypted with $K_{TB}$ and $K_{AB}$, while Alice knows nothing about $K_{TB}$. The results show that one adversarial user cannot obtain other parties’ secret information.

\subsubsection{The attack from adversarial TP}
In our protocol, TP is assumed to be adversarial and has full quantum capability. More importantly, all quantum resources and complex quantum state measurements can only be accomplished by TP. In fact, TP' s attacks pose the greatest threat to protocol security because she can take all possible attacks to steal the useful information, including preparing fake quantum states. Recall the steps of our protocol, Alice's (Bob's) secret information is encrypted by $K_{AB}$ and $K_{TA}$ ($K_{AB}$ and $K_{TB}$), where the value of $K_{AB}$ is unknown to TP. To obtain the secret of Alice (Bob), TP must get $K_{AB}$.  There are two kinds of attacks that TP may use to obtain $K_{AB}$. 

For the first kind of attack, TP may measure the received qubits in $Z$ basis instead of the Bell basis in step 5, and then TP announces  her measurement result in the form of $|\phi^+\rangle$ or $|\phi^-\rangle$ randomly. In this way, TP can obtain Alice and Bob's measurement results on the qubits that they choose the \textbf{\emph{measure}} operation. However, this kind of attack will destroy the state of $|\phi^+\rangle$. For the qubits that Alice and Bob both choose the \textbf{\emph{reflect}} operation, TP will have a $\frac{1}{2}$ probability to publish the wrong measurement result $|\phi^-\rangle$. Thus, this kind of attack will be detected by Alice and Bob.

For the second kind of attack, TP may prepare fake particles with $Z$ basis instead of the Bell state $|\phi^+\rangle$, and then she sends these fake particles to Alice and Bob. After that, TP can measure the received qubits with $Z$ basis to obtain Alice and Bob's measurement results. Following the protocol steps, TP is required to post the measurement result in the form of Bell state in step 5. Since TP prepares dummy particles, she can only publish the measurement results randomly as $|\phi^+\rangle$ or $|\phi^-\rangle$. Therefore, this kind of attack will be detected by Alice and Bob in step 5.

In addition, TP can obtain $K_{TA}$, $K_{TB}$, $Q_A$, $Q_B$ and the comparison result, in the course of the protocol. However, it is still helpless for TP to obtain Alice and Bob's secret, since she knows nothing about $K_{AB}$. The results show that the adversarial TP cannot obtain the classical users' privacy data.

\section{ Simulation based on IBM's Qiskit}
Here, we conduct an experimental simulation of the proposed SQPC protocol using IBM's Qiskit, aimed at demonstrating its feasibility and correctness. For the sake of simplicity, we will not consider any eavesdropping or attacks in the following procedures. 

\subsection{Simulation for the different Bell states}
In our protocol, Bell states serve as the fundamental information carrier, and thus, we first start with the preparation of Bell states and their circuit simulations. The corresponding simulation results are presented in Fig. 3. For example, to create the Bell state $|\phi^+\rangle$, we can use two qubits initially in the state $|0\rangle$, corresponding to $q_0$ and $q_1$ in Fig. 3a, and then perform a Hadamard gate (i.e., H gate) followed by a controlled-NOT gate (i.e., CNOT gate). To perform a Bell-based measurement on this state, we only need to apply the CNOT and H gates once, followed by a measurement. The measurement result obtained is always $|00\rangle$, corresponding to 00 in Fig. 3b. For $|\phi^-\rangle$, $|\psi^+\rangle$, and $|\psi^-\rangle$ their results by Bell-based measurements correspond to $01$, $10$, and $11$ in Fig. 3b, respectively.
\begin{figure}[ht]
  \centering
  \begin{subfigure}[b]{0.55\linewidth}
    \includegraphics[width=\linewidth,height=4.5cm]{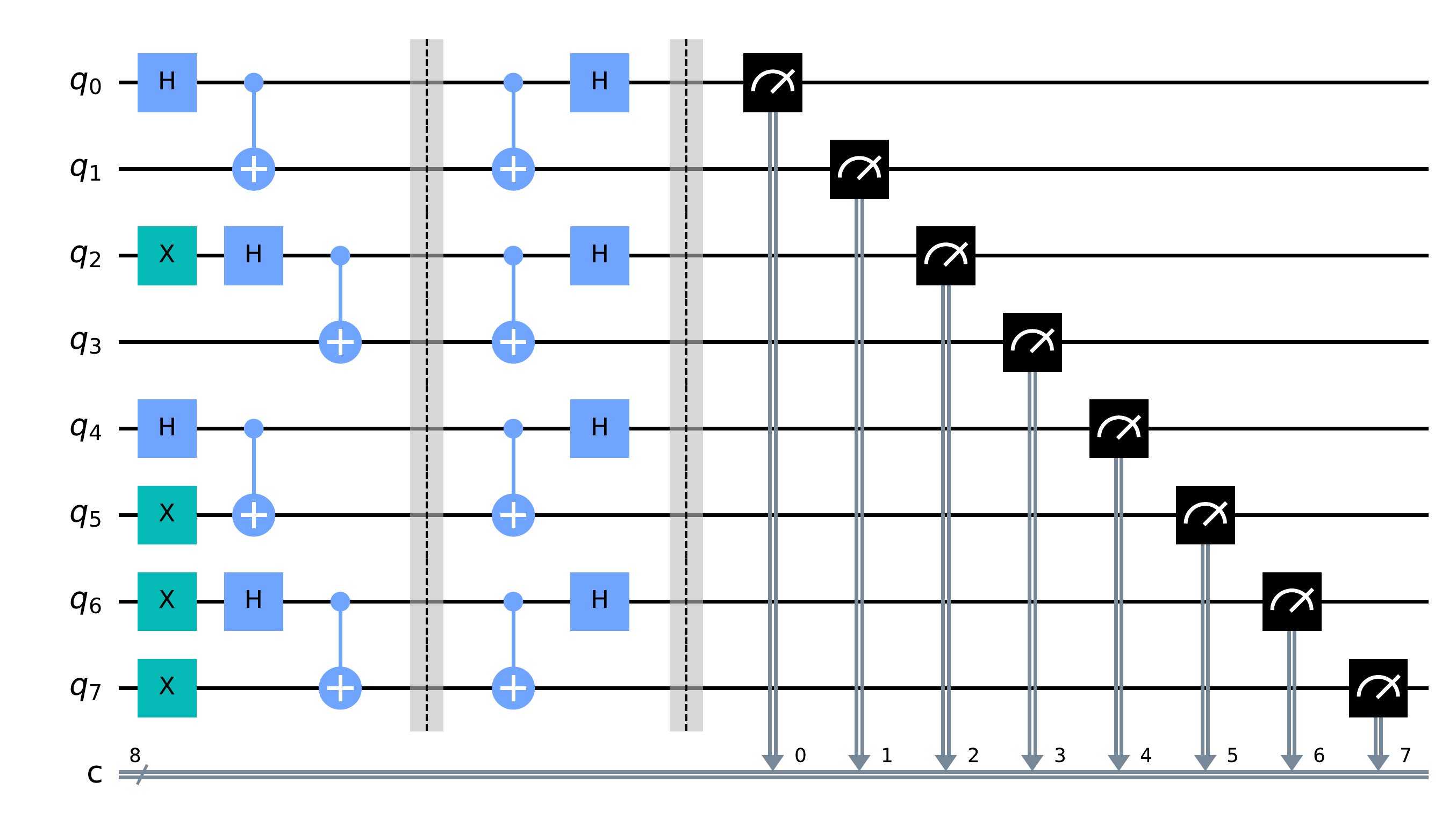}
    \caption{Preparation and measurement of four different Bell states (i.e., $|\phi^+\rangle$,$|\phi^-\rangle$,$|\psi^+\rangle$,$|\psi^-\rangle$)}
    \label{fig:sub1}
  \end{subfigure}
  \hspace{0.05\linewidth}
  \begin{subfigure}[b]{0.35\linewidth}
    \includegraphics[width=\linewidth,height=4.3cm]{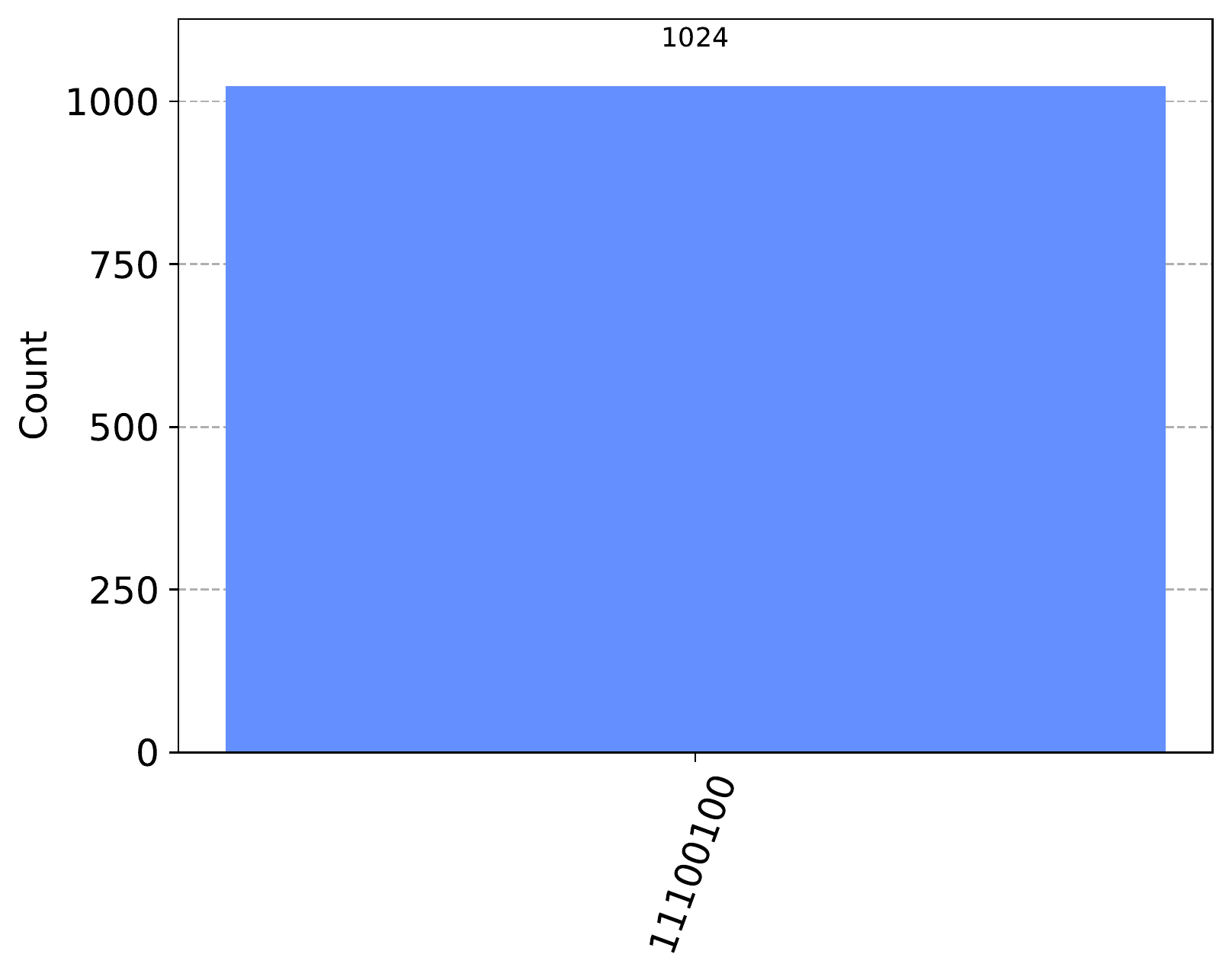}
    \caption{The results of the Bell-based measurement for the different Bell states}
    \label{fig:sub2}
  \end{subfigure}
  \caption{Four different Bell state preparation methods and corresponding circuit simulations with Bell-based measurements. By setting different initial states, namely $|00\rangle$, $|01\rangle$, $|10\rangle$, $|11\rangle$, the preparation and measurement of Bell states $|\phi^+\rangle$,$|\phi^-\rangle$,$|\psi^+\rangle$,$|\psi^-\rangle$ can be realized by using H gate and CNOT gate.}
  \label{fig:3}
\end{figure}

\subsection{Simulation for the proposed SQPC protocol.}
According to the protocol description, TP will take the two Bell states $|\phi^+\rangle|\phi^+\rangle$ as a group and randomly perform entanglement swapping, and finally send them to Alice. After Alice and Bob complete their operations, TP will perform corresponding operations on the received qubits to establish the secret key and conduct security detection. As described in step 5, Alice and Bob will randomly select half of the Bell state groups, and then let TP measure these qubits in Bell basis and publish the measurement results. 

\begin{figure}[!ht]
  \centering
  \begin{subfigure}[b]{0.55\linewidth}
    \includegraphics[width=\linewidth,height=3.7cm]{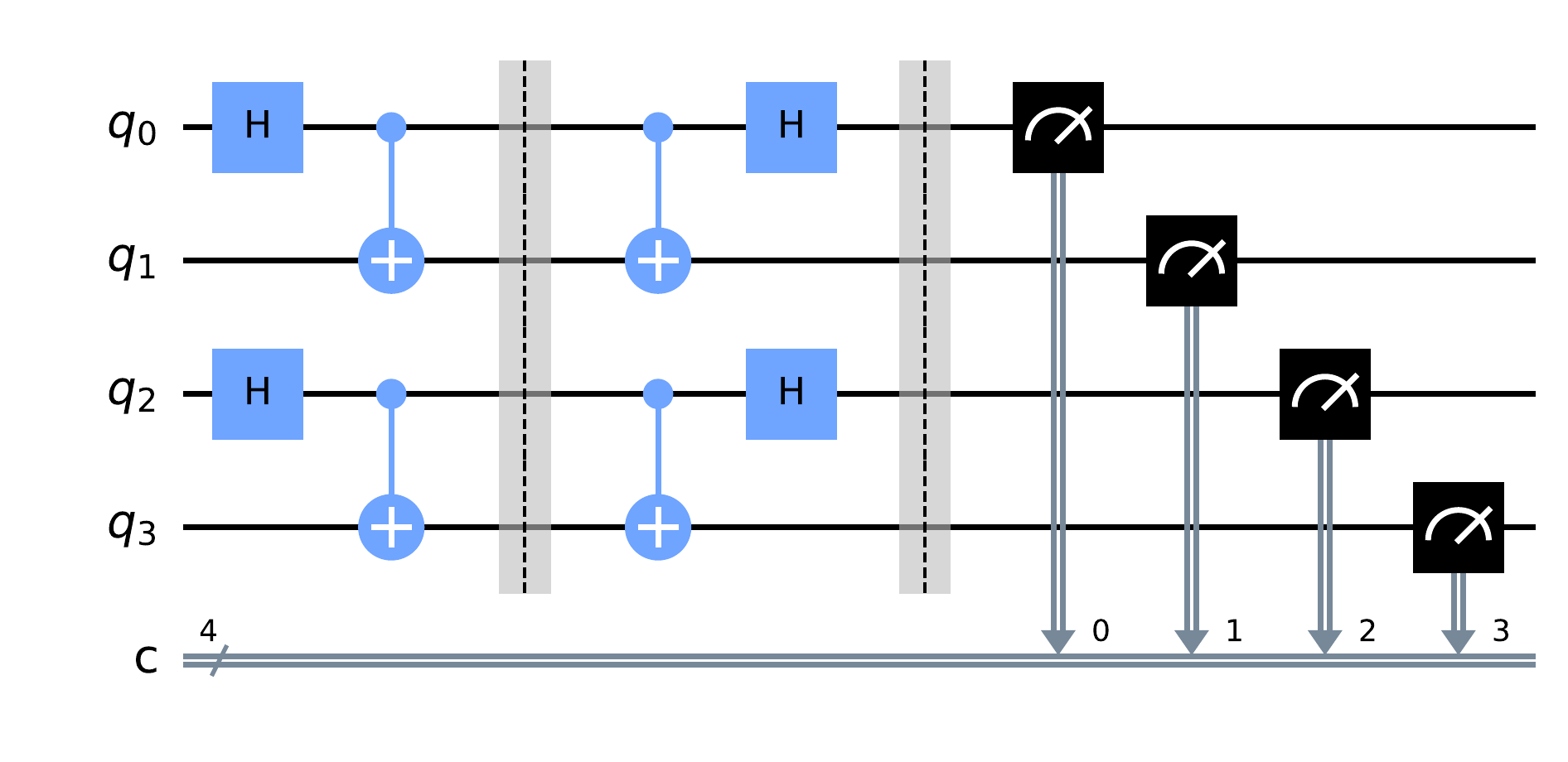}
    \caption{Two Bell states as a group without entanglement swapping.}
    \label{fig:sub1}
  \end{subfigure}
  \hspace{0.05\linewidth}
  \begin{subfigure}[b]{0.35\linewidth}
    \includegraphics[width=\linewidth,height=3.5cm]{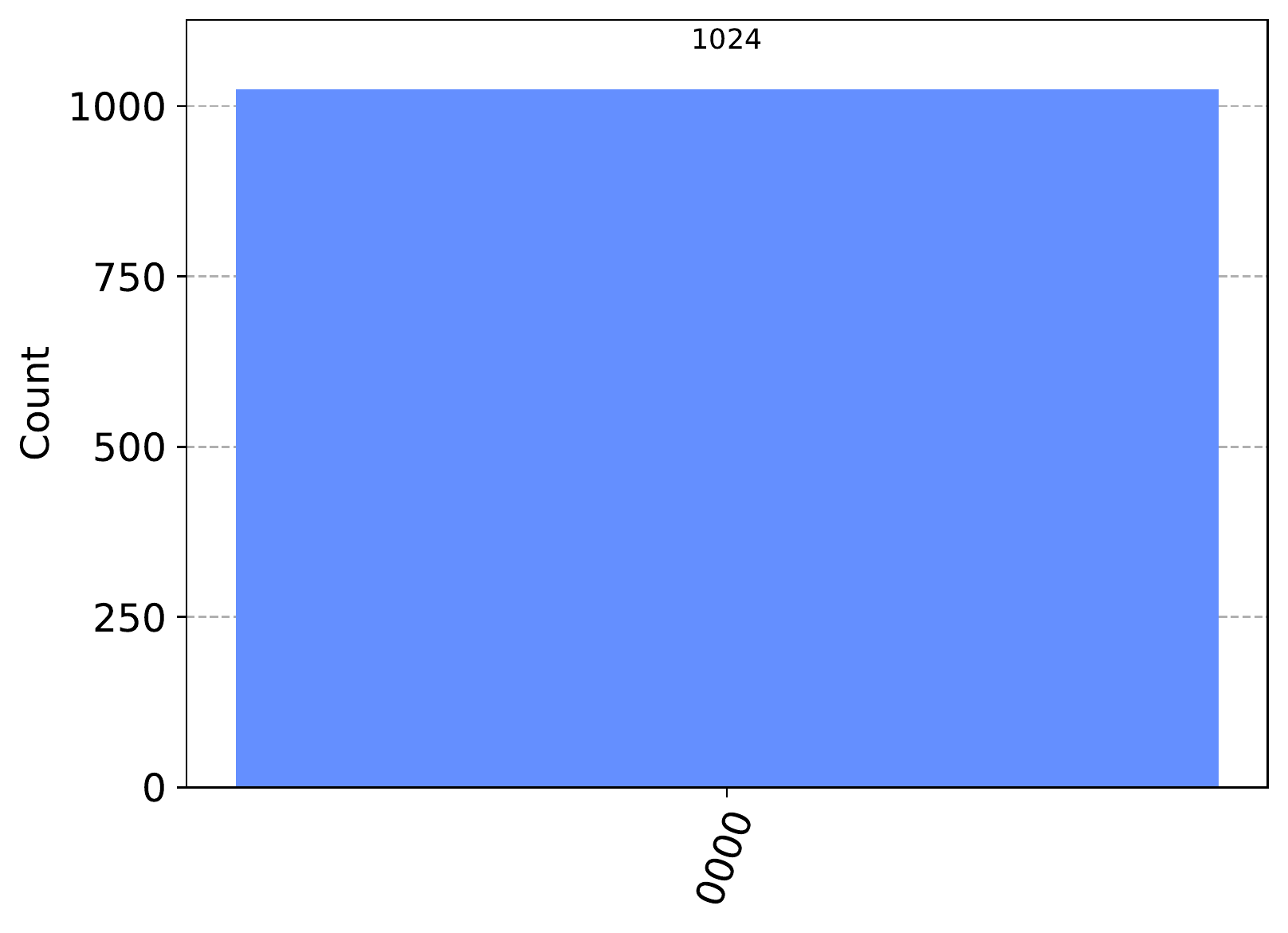}
    \caption{The simulation results of (a).}
    \label{fig:sub2}
  \end{subfigure}
  \hspace{0.05\linewidth}
    \begin{subfigure}[b]{0.55\linewidth}
    \includegraphics[width=\linewidth,height=3.7cm]{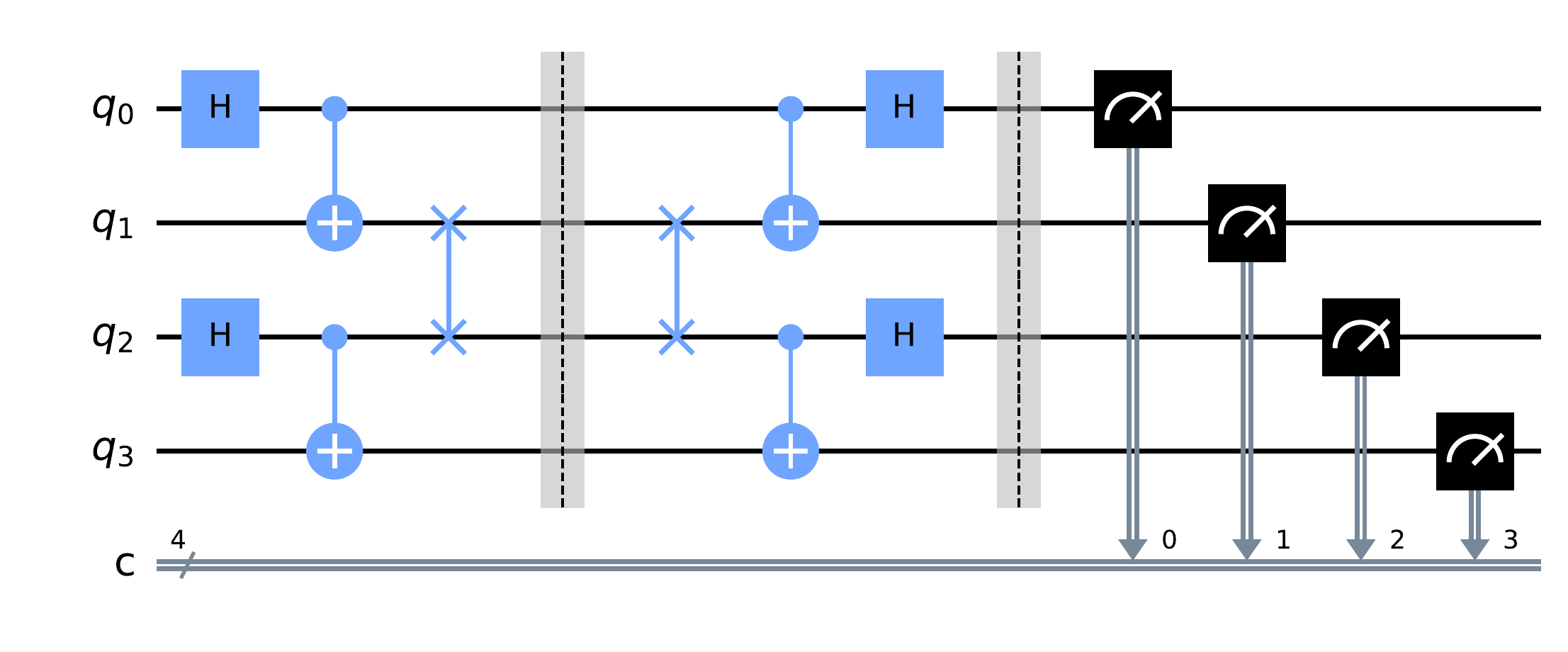}
    \caption{Two Bell states as a group with entanglement swapping.}
    \label{fig:sub3}
  \end{subfigure}
  \hspace{0.05\linewidth}
  \begin{subfigure}[b]{0.35\linewidth}
    \includegraphics[width=\linewidth,height=3.5cm]{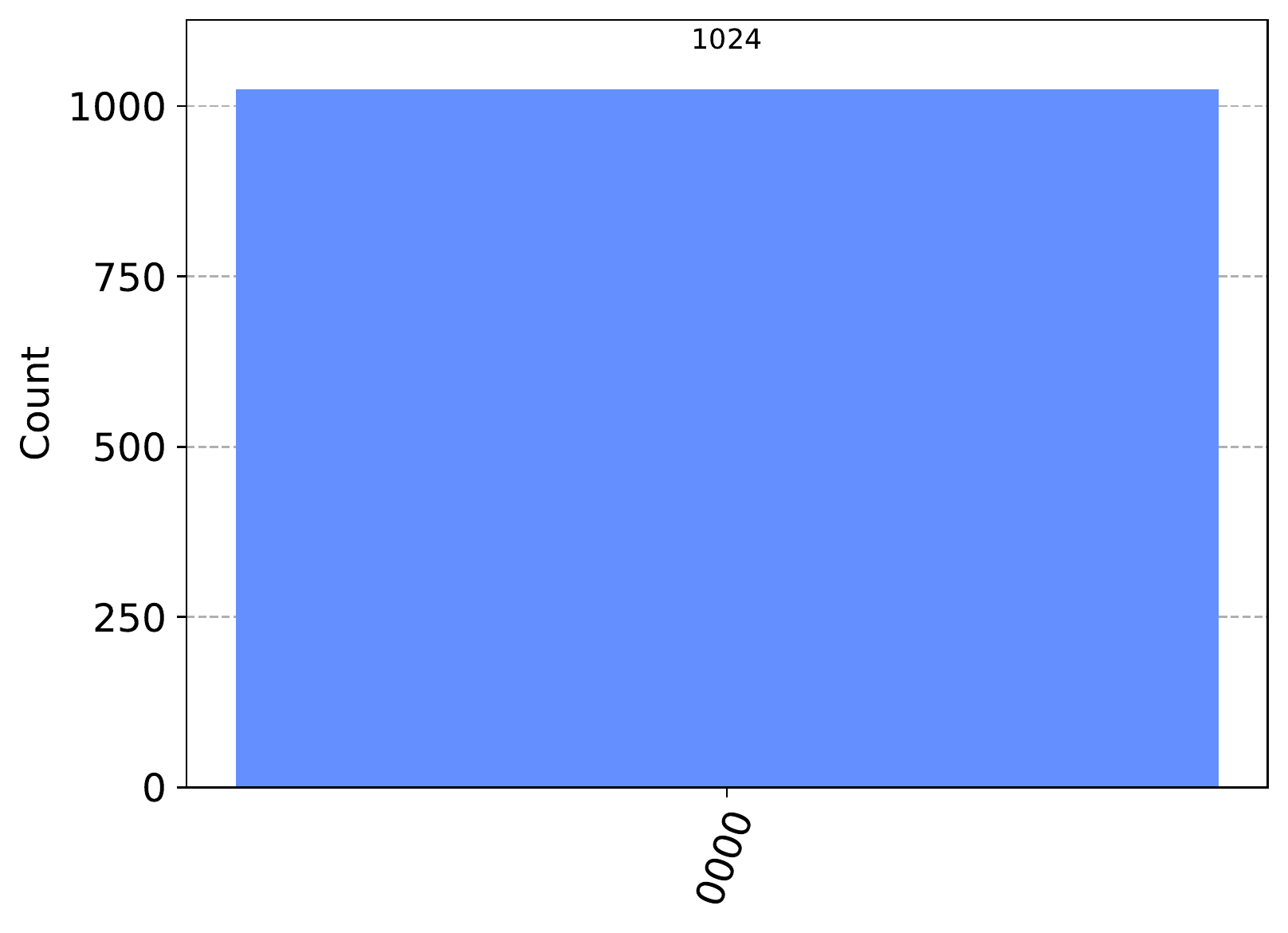}
    \caption{The simulation results of (c).}
    \label{fig:sub4}
  \end{subfigure}
  \caption{Circuits and simulation results for both Alice and Bob performing \textbf{\emph{reflect}} operation. }
  \label{fig:4}
\end{figure}

Here, we first focus on the case where both Alice and Bob choose the \textbf{\emph{reflect}} operation. The corresponding quantum circuits and simulation results are shown in Fig. 4. As illustrated in Fig. 4a, no entanglement swapping occurs between the two Bell states, and TP ultimately performs Bell-based measurements on the received qubits. In contrast, Fig. 4c shows that the two Bell states undergo entanglement swapping, and TP makes the entanglement swapping on the received qubits to restore the initial Bell state before conducting Bell-based measurements. In this case, the results obtained by TP should all be $|\phi^+\rangle|\phi^+\rangle$ (i.e., the simulation result should always  be 0000), if no eavesdropper is present. From the Fig. 4b and Fig. 4d, it can be concluded that TP's measurement results are always at $0000$. Therefore, the simulation result is consistent with the designed protocol.

Next, we focus on the case that Alice and Bob both choose the \textbf{\emph{measure}} operation on the same qubits, and TP measures these qubits in Bell basis. In this case, Alice and Bob can securely generate a secret key sequence $K_{AB}$ without any information leakage to TP.
\begin{figure}[!ht]
  \centering
  \begin{subfigure}[b]{0.53\linewidth}
    \includegraphics[width=\linewidth,height=4.1cm]{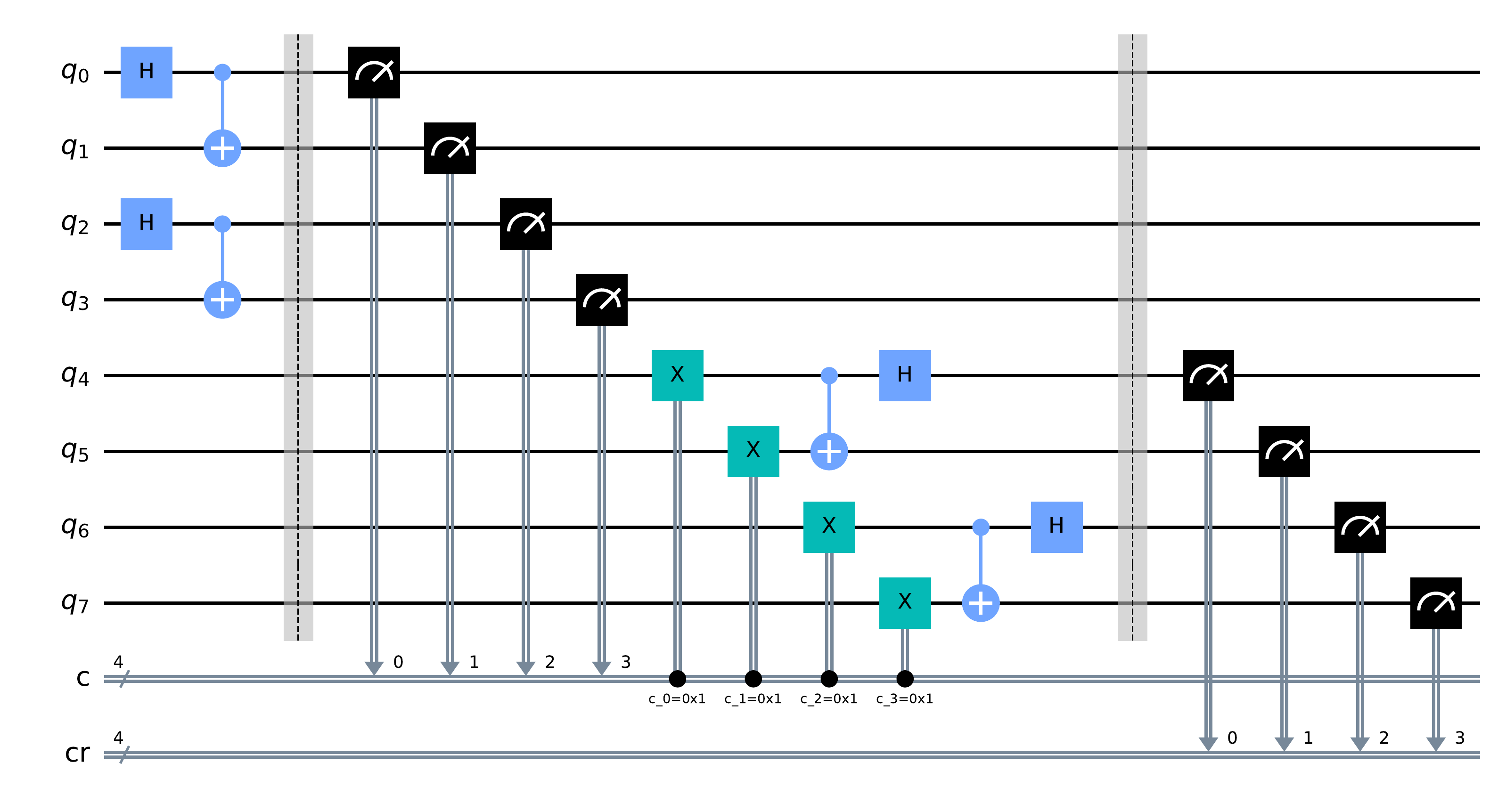}
    \caption{Measurement on two unentangled-swapped Bell states.}
    \label{fig:sub1}
  \end{subfigure}
  \hspace{0.05\linewidth}
  \begin{subfigure}[b]{0.37\linewidth}
    \includegraphics[width=\linewidth,height=3.8cm]{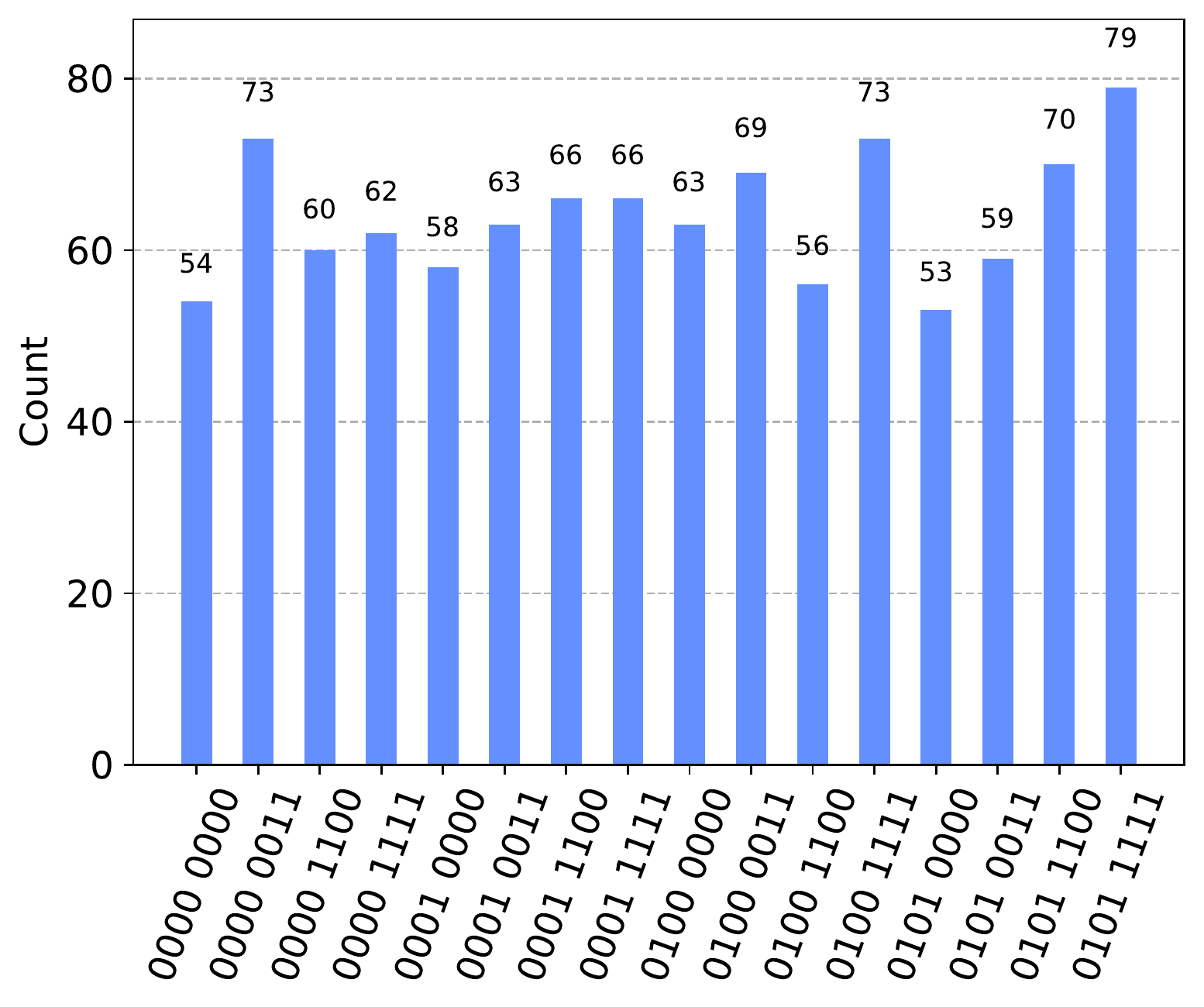}
    \caption{The simulation results of (a).}
    \label{fig:sub2}
  \end{subfigure}
  \hspace{0.05\linewidth}
    \begin{subfigure}[b]{0.53\linewidth}
    \includegraphics[width=\linewidth,height=4.1cm]{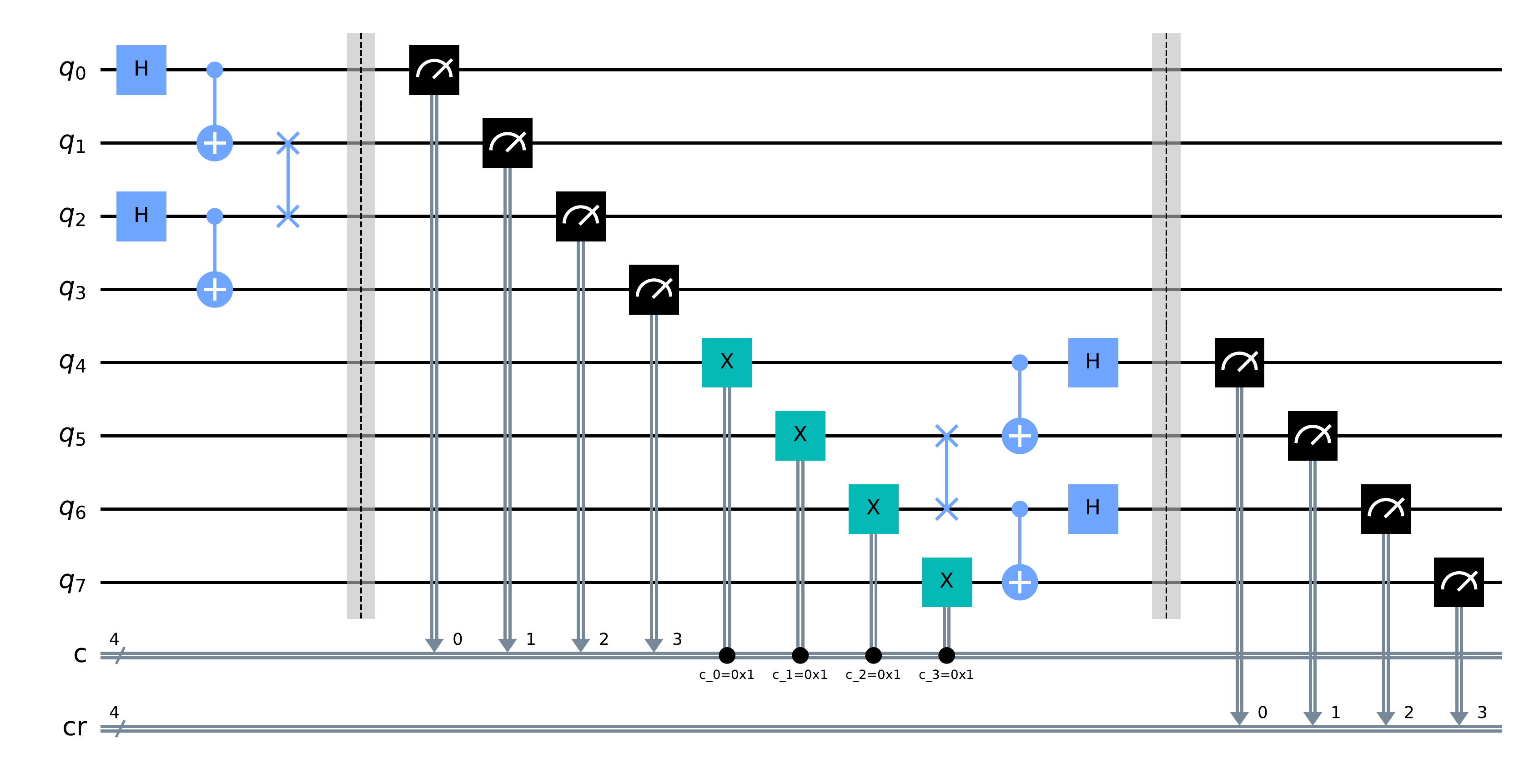}
    \caption{Measurement on two entangled-swapped Bell states.}
    \label{fig:sub3}
  \end{subfigure}
  \hspace{0.05\linewidth}
  \begin{subfigure}[b]{0.37\linewidth}
    \includegraphics[width=\linewidth,height=3.8cm]{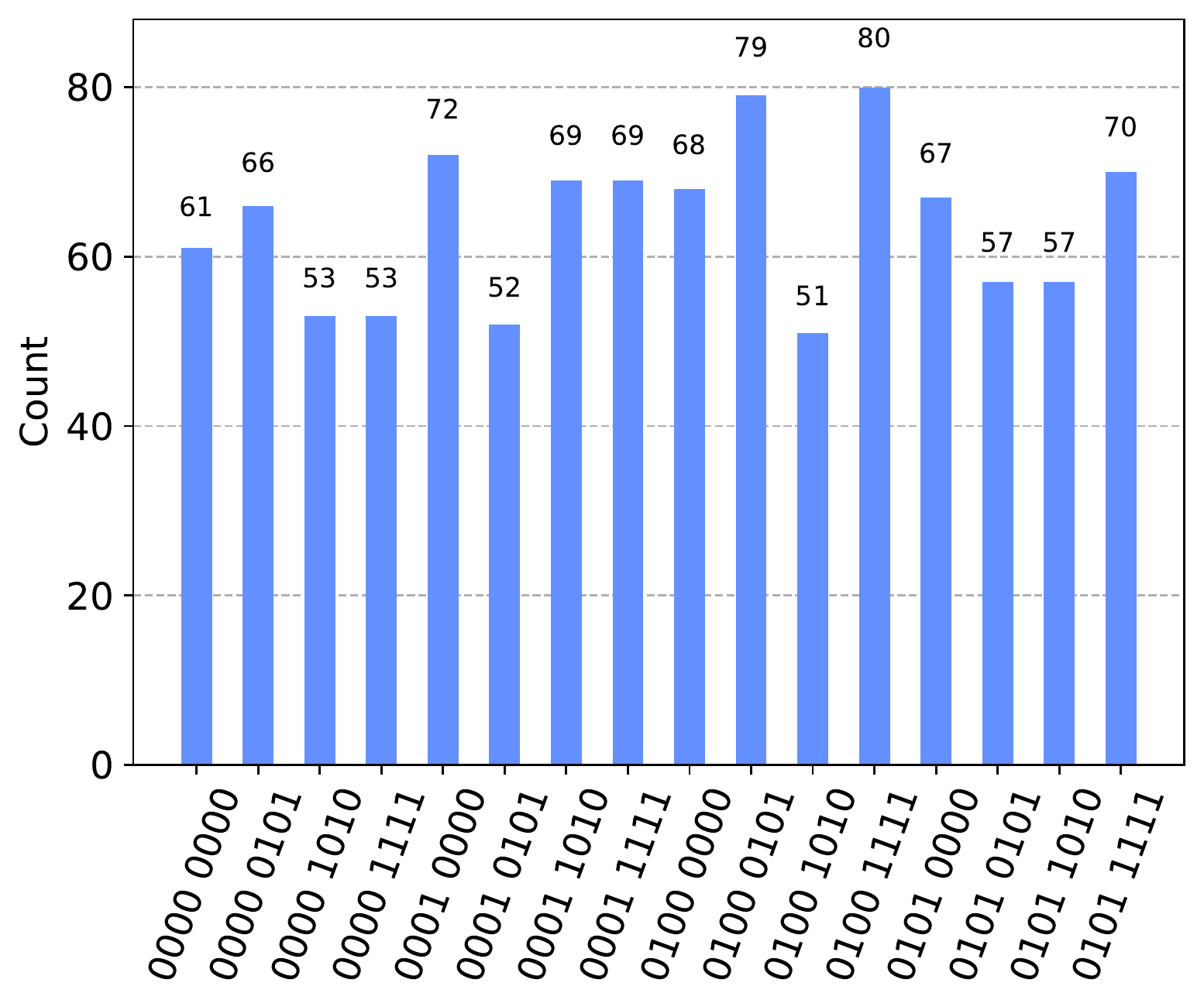}
    \caption{The simulation results of (c).}
    \label{fig:sub4}
  \end{subfigure}
  \caption{Quantum circuits and simulation results for the case where Alice and Bob measure all four qubits of two Bell states and TP measures in Bell basis. Classical registers $c$ and $cr$ record the measurement results of Alice and Bob, and TP, respectively. Here, $(q_4, q_5, q_6, q_7)$ represents the qubits remade by Alice and Bob. }
  \label{fig:5}
\end{figure}

For example, assume that Alice and Bob simultaneously perform \textbf{\emph{measure}} operation on all four qubits of two Bell states, while TP measures these qubits in Bell basis. The corresponding quantum circuits are presented in Fig. 5. It is worth noting that Alice and Bob can generate secret keys as long as they both perform measurement operations on at least one qubit, and TP performs Bell-based measurements on the corresponding qubits. From Fig. 5, it is not difficult to see that the measurement results of TP are distributed in the four results of 0000, 0001, 0100, and 0101, corresponding to $|\phi^+\rangle|\phi^+\rangle$, $|\phi^+\rangle|\phi^-\rangle$, $|\phi^-\rangle|\phi^+\rangle$, and $|\phi^-\rangle|\phi^-\rangle$, respectively. For each specific measurement of TP, Alice and Bob's measurements are in four different outcomes, regardless of whether entanglement swapping occurs between the two Bell states. Therefore, TP is unable to deduce the measurement results of Alice and Bob based on her measurements, thereby TP knows nothing about the secret key $K_{AB}$ shared by Alice and Bob. From the example given above, we can see that the simulation results meet the requirements of the designed protocol.

After completing step 5, Alice and Bob will publicly disclose the operations they have performed on the remaining qubits. TP will then conduct different measurement operations on these qubits to establish the secret keys $K_{TA}$, $K_{TB}$, and perform security detection. As described in step 6, there will be three cases to consider: 1) Both Alice and Bob choose the \textbf{\emph{reflect}} operation on the received qubits directly; 2) Alice and Bob choose different operation on the received qubits; 3) Both Alice and Bob select the \textbf{\emph{measure}} operation for the received qubits.

For the first case, it is used to eavesdropping checking, and its quantum circuits and simulation results can be referred to Fig. 4. In the following, attention is focused on the other two cases, which are essential to establishing the secret key and ensuring secure communication. 

Assume that Alice performs \textbf{\emph{measure}} operation on the first and third qubits of two Bell states while the other qubits undergo \textbf{\emph{reflect}} operation, and Bob performs \textbf{\emph{measure}} operation on the first and second qubits of the same group while the other qubits undergo \textbf{\emph{reflect}} operation. This scenario is of particular interest because it involves situations where Alice and Bob perform different operations on the same qubit, as well as situations where they both measure the same qubit. We also use IBM's qiskit to give the quantum circuits and simulation results of the above situations, as shown in Fig. 6.
\begin{figure}[!ht]
  \centering
  \begin{subfigure}[b]{0.55\linewidth}
    \includegraphics[width=\linewidth,height=4.2cm]{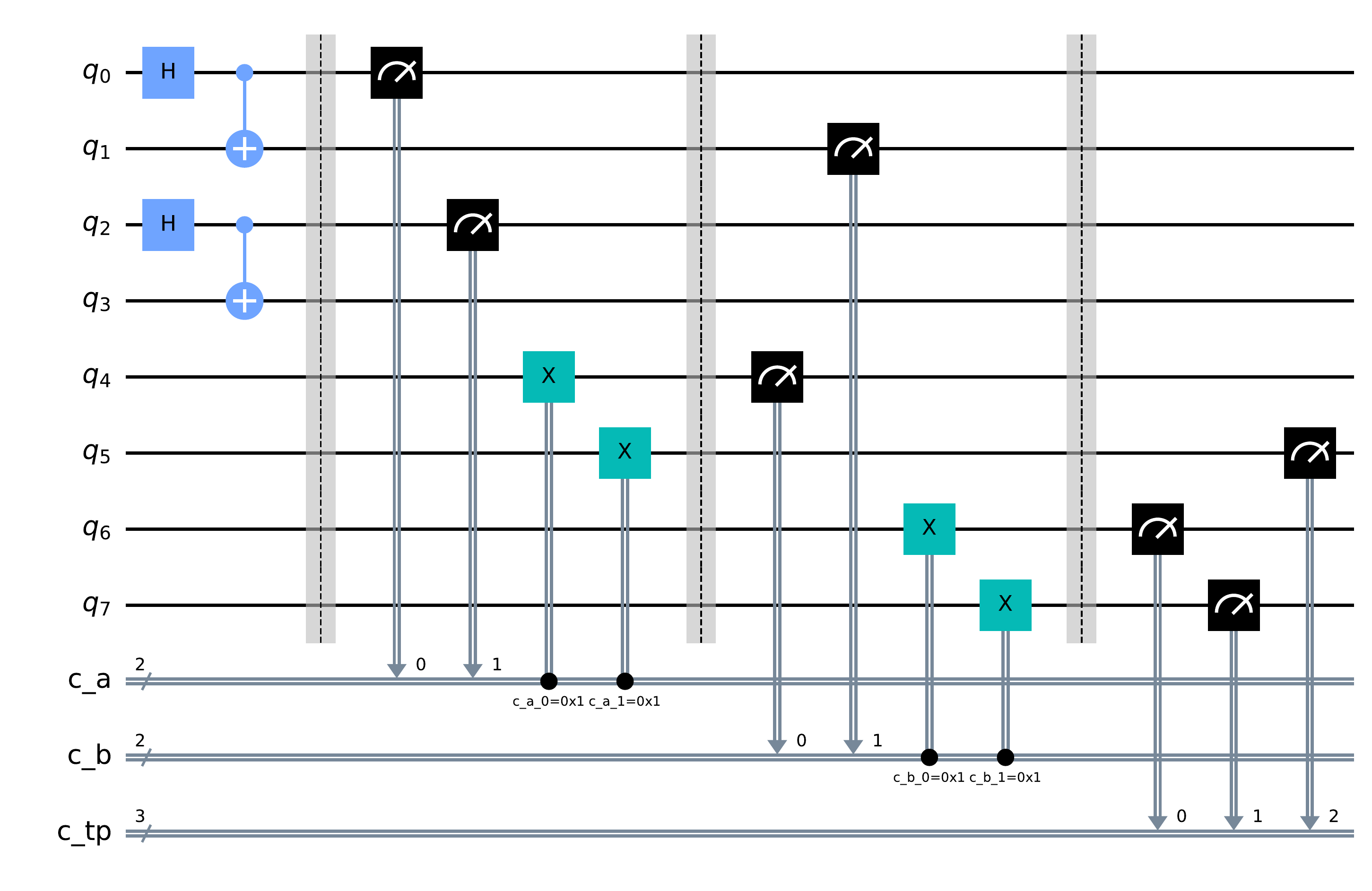}
    \caption{Case of two unentangled-swapped Bell states.}
    \label{fig:sub1}
  \end{subfigure}
  \hspace{0.05\linewidth}
  \begin{subfigure}[b]{0.35\linewidth}
    \includegraphics[width=\linewidth,height=4cm]{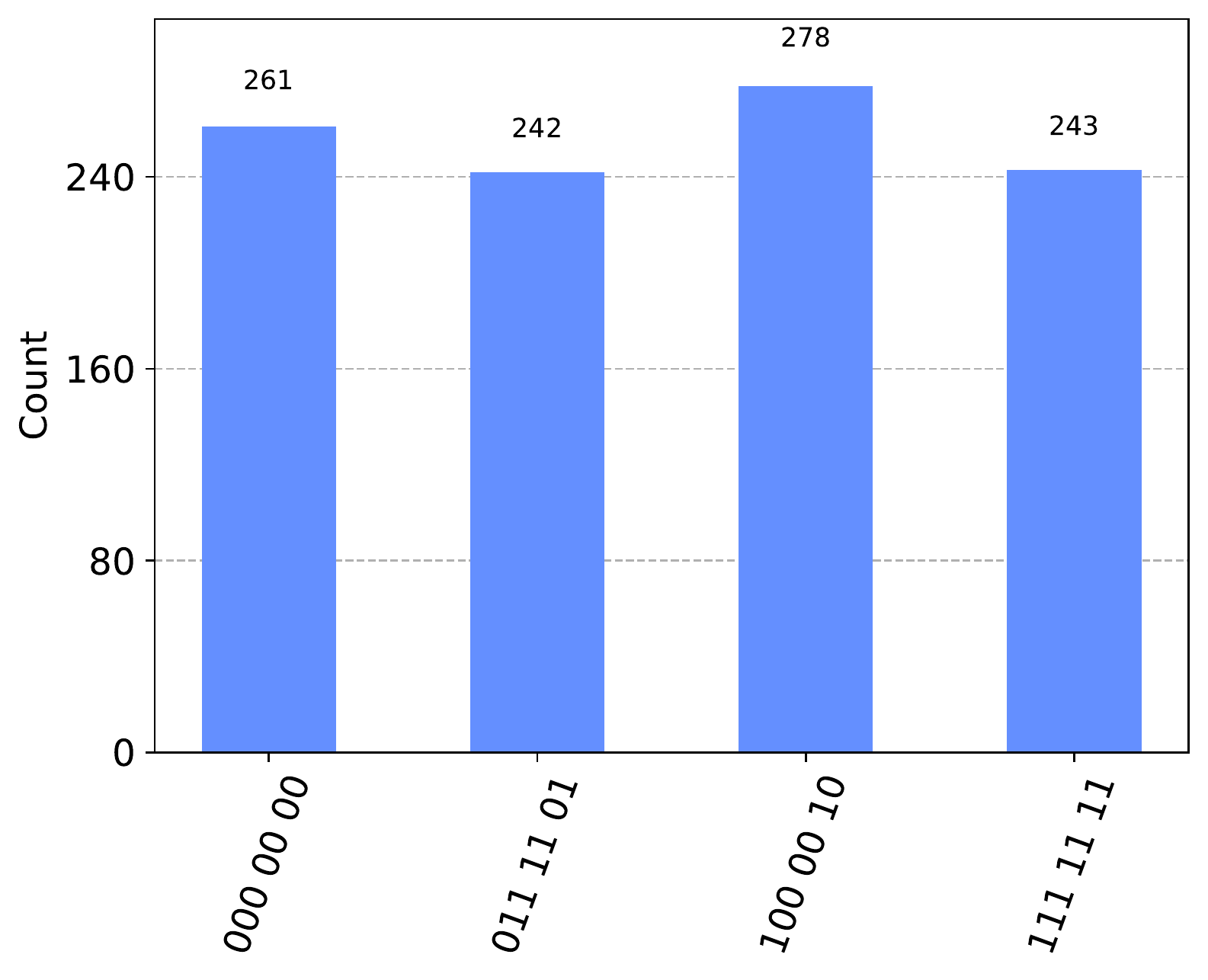}
    \caption{The simulation results of (a).}
    \label{fig:sub2}
  \end{subfigure}
  \hspace{0.05\linewidth}
    \begin{subfigure}[b]{0.55\linewidth}
    \includegraphics[width=\linewidth,height=4.2cm]{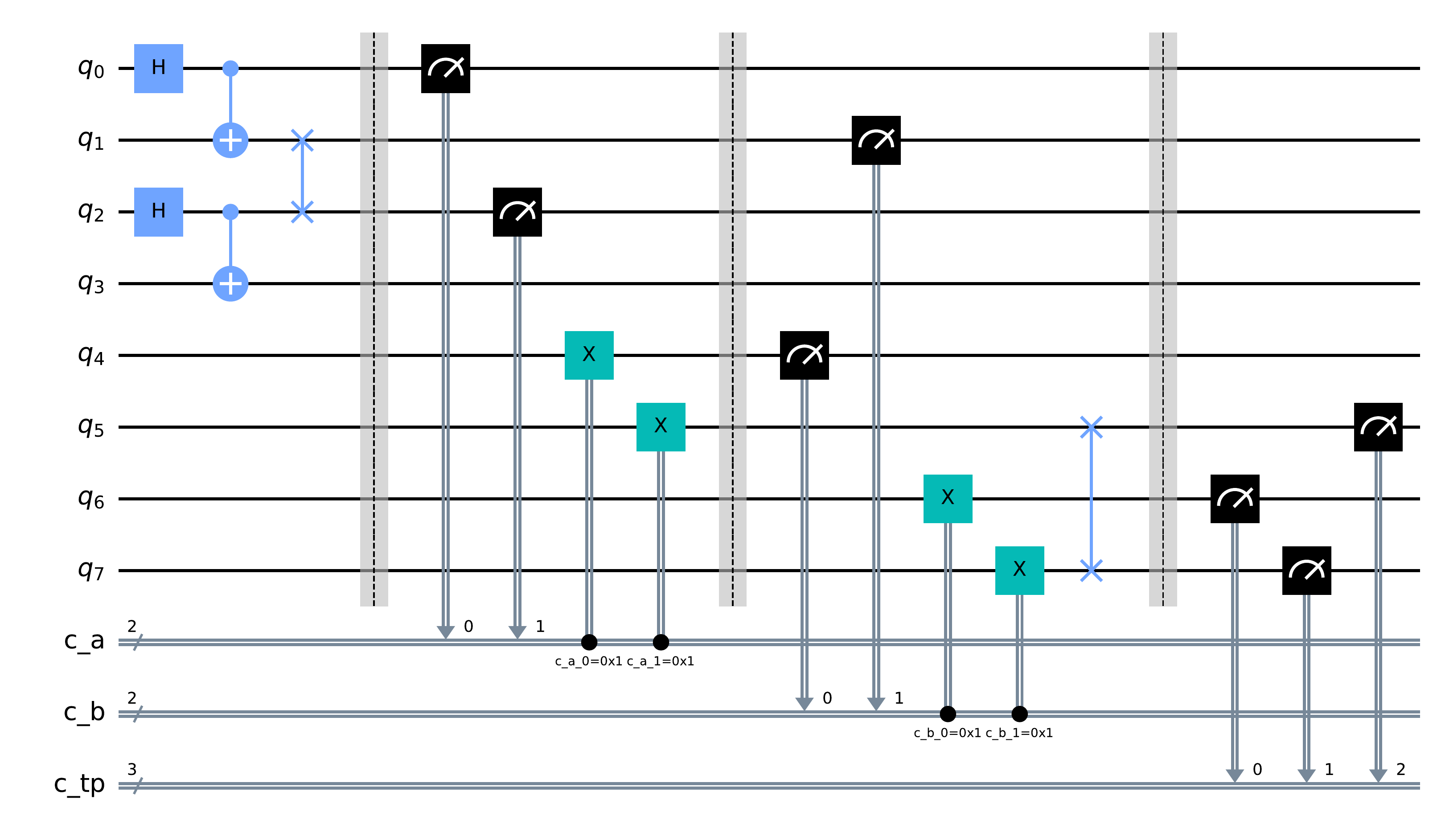}
    \caption{Case of two entanglement-swapped Bell states.}
    \label{fig:sub3}
  \end{subfigure}
  \hspace{0.05\linewidth}
  \begin{subfigure}[b]{0.35\linewidth}
    \includegraphics[width=\linewidth,height=4cm]{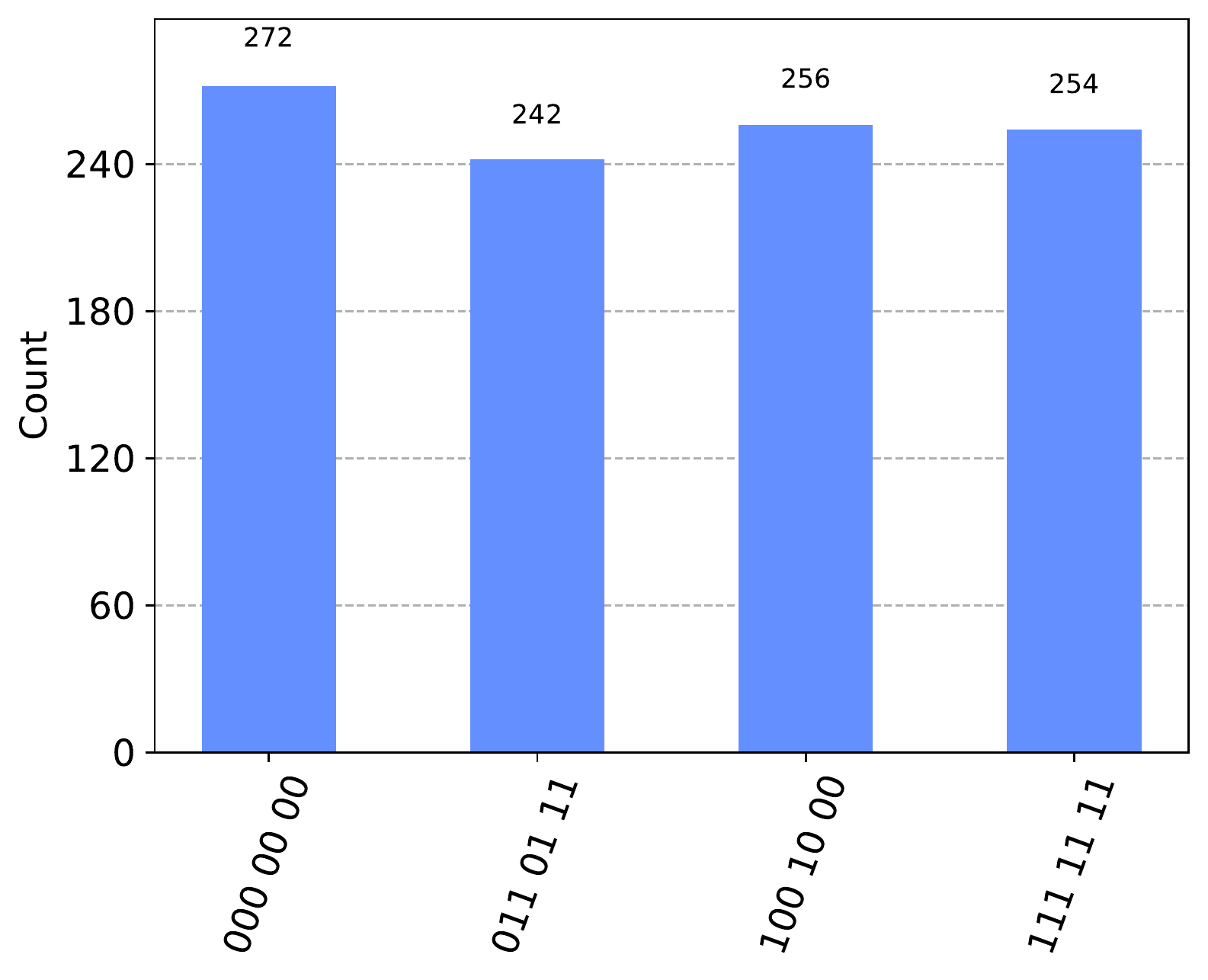}
    \caption{The simulation results of (c).}
    \label{fig:sub4}
  \end{subfigure}
  \caption{Quantum circuits and simulation results for the situation that Alice performs \textbf{\emph{measure}} operation on the first and third qubits of Bell state group while Bob performs \textbf{\emph{measure}} operation on the first and second qubits ot the same group.
 Classical registers $c\_a$, $c\_b$, and $c\_tp$ record the measurement results of Alice, Bob, and TP, respectively. Note that after Alice and Bob measure the received qubits, they will prepare new qubits based on the measurement results. Here, we use $(q_4, q_5)$ and $(q_6, q_7)$ to represent the qubits remade by Alice and Bob.}
  \label{fig:6}
\end{figure}

We will now provide a detailed description of Fig. 6. Specifically, in Fig. 6a, entanglement swapping is not performed by TP for the two Bell states. Hence, the classical register $c\_a$ records the measurement outcomes of the first and third qubits, while $c\_b$ records the results of the first and second qubits. In addition, for register $c\_tp$, it records the measurement results of the first three qubits. For instance, ``011 11 01'' in the second column of Fig. 6b represents that the measurement results recorded by  $c\_tp$, $c\_b$, and $c\_a$ are ``011'', ``11'', and ``01'', respectively. Please note that the leftmost bit in register corresponds to the measurement result of highest qubit. Hence, ``011'' in register $c\_tp$ represents the measurements of the third, second, and first qubits, respectively. ``11'' in register $c\_b$ denotes the measurements of the second and first qubits. ``01'' in register $c\_a$ represents the measurements of the third and first qubits. It is easy to see that Alice and TP have the same measurement result for the third qubit, while Bob and TP have the same measurement result for the second qubit. Thus the measurement results of the third and second qubit can be used as the key of $K_{TA}$ and $K_{TB}$ respectively. For the first qubit, Alice, Bob, and TP all measure it, and the measurement results are the same. This condition also serves as a check against potential eavesdropping.

Furthermore, for the Fig. 6c and Fig. 6d, the analysis process is similar to the above. It is worth noting that in this case, the entanglement swapping occurs between the second and third qubits, which means that the qubits measured by Alice correspond to the first and second qubits, whereas the qubits measured by Bob correspond to the first and third qubits. From the Fig. 6d, we thus can obtain the same result that TP and Alice have the same measurement result on the second qubit, while TP and Bob have the same measurement result on the third qubit. Therefore, the simulation results shown in Fig. 6 are consistent with the results required by the protocol. That is, TP and Alice can establish the secure key $K_{TA}$, while TP and Bob can establish the secure key $K_{TB}$. 

Putting everything together, we verify that the designed protocol satisfies the feasibility and correctness that Alice and Bob, TP and Alice, and TP and Bob can establish keys with each other. This also shows that our SQPC protocol is feasible and correct.

\section{ Comparison and discussion}
\label{sec:5}

In this paper, we propose a semi-quantum private comparison protocol based on entanglement swapping. To better highlight the characteristics of the protocol, we compare the proposed protocol with previous related work. As qubit efficiiency is an important metric for evaluating the performance of semi-quantum protocols, it has always been a focus of reserach. Therefore, in the followsing part, we will first analyze the qubit efficiency of our SQPC protocol.

Qubit efficiency is defined as $\eta=\frac{c}{q+b}$, where $c$, $q$, and $b$ are the number of shared classical bits, the number of consumed qubits, and the number of classical bits needed, respectively \cite{38}. In our protocol, Alice and Bob each have $n$ secret bits, resulting $c=n$. To complete the whole protocol, TP needs to generate $4n$ Bell states (i.e., $8n$ qubits), while Alice and Bob need to prepare $4n$ qubits each as a replacement for their measured qubits. Hence, the total number of consumed qubits is $16n$. Moreover, Alice and Bob need $2n$ bits to publish their encrypted data, and TP requires 1 bit to announce the comparison result, resulting $b=2n+1$. Therefore, the qubit efficiency of our protocol is $\frac{n}{18n+1}$. We can similarly calculate the qubit efficiency of Refs. \cite{23,24,25,26,27,28,29} and the specific results are shown in Table 2. Note that in calculating the protocol qubit efficiency, we take into account the additional overhead of establishing pre-shared key. It is evident from the table that our protocol has an advantage in qubit efficiency compared to those using distributed transmission. 
Compared to Ref. \cite{29} that also adopts circular transmission, our protocol has the same level of qubit efficiency.
However, it is worth noting that our protocol employs Bell states and entanglement swapping, whereas Ref. \cite{29} relies on single particles. 
This distinction also reflects that with our approach, even with the use of entanglement swapping, efficient SQPC protocol can be achieved.

Next, we discuss the the role of entanglement swapping in the proposed protocol. During the protocol, two Bell states are transmitted as a group between TP, Alice, and Bob, where the qubits of each group, randomly, undergo entanglement swapping. The entanglement correlation between qubits in each Bell state is also used to verify the honesty of TP in preparing the initial quantum resource. Our approach guarantees the security of protocol, because the attacker cannot determine whether each group has undergone entanglement swapping, so that he cannot effectively steal helpful information. Compared with previous SQPC protocol based on entanglement swapping \cite{23}, our protocol make it possible for each qubit to contribute keys by using the nature of entanglement swapping and circular transmission, which greatly improves the efficiency of the protocol.

We then turn our attention to the complexity and scalability of the protocol. Unlike previous protocols \cite{24,26,27,28}, the proposed protocol does not require an additional semi-quantum key distribution protocol to establish secret key between classical users in advance, thus reducing the complexity of the protocol. Moreover, the circular transmission employed in our protocol provides good scalability and a possible way for achieving multi-party privacy comparison. Specifically, suppose more than two classical users want to compare the equivalence of their secret data. In that case, it is simply a matter of adding new users between Bob and TP and establishing quantum state transmission between each other. 

Overall, our protocol achieves high qubit efficiency and does not require pre-shared keys. Although our protocol utilizes the entanglement swapping with Bell states, the protocol's efficiency is not inferior to those using single-particle states or not using entanglement swapping. Thus, our SQPC protocol may have wider application scenarios.
\begin{table}[htp]
\centering 
\tabcolsep 4pt
\caption{Comparison between our SQPC protocol and previous ones }
\label{tab:1}       
\begin{tabular}{p{0.08\linewidth}<{\centering}p{0.1\linewidth}<{\centering}p{0.12\linewidth}<{\centering}p{0.11\linewidth}<{\centering}p{0.11\linewidth}<{\centering}p{0.11\linewidth}<{\centering}p{0.1\linewidth}<{\centering}p{0.09\linewidth}<{\centering}}
\hline
 \noalign{\smallskip}
&Quantum resource  & Transmission mode  & Entanglement swapping &  usage of pre-shared key  & Pre-shared key cost & Comparison cost & Qubit efficiency\\   
  \hline
  \noalign{\smallskip}
  Ref.\cite{23} & Bell states  & Distributed                    & \checkmark  & \ding{53}  & $0$    & $162n+1$    & $\frac{n}{162n+1}$\\
  \noalign{\smallskip}
  Ref.\cite{24} & Two-particle product particles  & Distributed & \ding{53} & \checkmark   & $16n$  & $44n+1$  & $\frac{n}{60n+1}$\\
  \noalign{\smallskip}
  Ref.\cite{25} & Single particles  & Distributed               & \ding{53}  & \ding{53}   & $0$    & $52n+1$    & $\frac{n}{52n+1}$\\
  \noalign{\smallskip}
  Ref.\cite{26} & Bell states  & Distributed                    & \ding{53} & \checkmark   & $40n$  & $60n+1$  & $\frac{n}{102n+1}$\\
  \noalign{\smallskip}
  Ref.\cite{27} & Bell states  & Distributed                    & \ding{53} & \checkmark   & $40n$  & $12n+1$  & $\frac{n}{52n+1}$\\ 
  \noalign{\smallskip}
  Ref.\cite{28} & Three-particles G-like states  & Distributed  & \ding{53} & \checkmark   & $40n$  & $13n+1$  & $\frac{n}{53n+1}$\\ 
  \noalign{\smallskip}
  Ref.\cite{29} & Single particles  & Circular                  & \ding{53} & \ding{53}    & $0$    & $18n+1$    & $\frac{n}{18n+1}$\\ 
  \noalign{\smallskip}
  Our protocol & Bell states  & Circular                        & \checkmark & \ding{53}   & $0$    & $18n+1$    & $\frac{n}{18n+1}$\\ 
\hline
\end{tabular}
\end{table}

\section{Conclusion }
\label{sec:6}

In this paper, we present a feasible SQPC protocol based on entanglement swapping of Bell states, where Bell states are randomly performed entanglement swapping and transmitted between TP, Alice, and Bob. The proposed SQPC protocol does not require an additional semi-quantum key distribution protocol to pre-share the key between two classical users, which greatly reduce the burden on quantum resources. By exploiting the properties of entanglement swapping and entanglement correlation, the security of our protocol is guaranteed, and attackers connot steal useful information without introducing errors. Compared with previous related work, our protocol has high qubit efficiency even with the use of entanglement swapping and Bell states. Our protocol combines entanglement swapping and circular transmission, making it possible for each qubit to act as a information carrier and be used for final encryption. This provides a possibility to design efficient semi-quantum cryptography protocols using entanglement swapping.

There are still many interesting questions for future research. In this work, we only consider ideal environments, while practical devices in semi-quantum environments are just beginning to be realized \cite{39,40}. Applying some of these techniques to the protocol we presented here would be interesting.

\section*{Acknowledgments} 
This work is partially supported by the National Natural Science Foundation of China under Grant 62271070, the BUPT Excellent Ph.D Students Foundation under Grant CX2021117, and the China Scholarship Council under Grant 202206470006.

\end{document}